\documentclass[11pt]{article}
\usepackage[margin=0.625in]{geometry}

\usepackage[T1]{fontenc}

\usepackage[style=science, doi=true, backend=biber]{biblatex}
\usepackage{graphicx}
\usepackage{float}
\PassOptionsToPackage{hyphens}{url}
\usepackage{bookmark}
\usepackage[all]{hypcap}
\usepackage[version=4]{mhchem}
\usepackage[capitalise]{cleveref}
\hypersetup{
    colorlinks=true,
    linkcolor=blue,
    filecolor=magenta,      
    urlcolor=cyan,
}

\usepackage{wrapfig}
\usepackage{sidecap}
\usepackage{subcaption}
\usepackage[labelfont={bf,footnotesize}, textfont=footnotesize, skip=8pt plus 2pt minus 2pt]{caption}

\usepackage{chngcntr}
\usepackage{appendix}

\usepackage{mathtools} %
\usepackage{amssymb}
\usepackage{amsfonts}
\usepackage{xspace}
\usepackage{changepage}

\usepackage{lineno} %

\usepackage{amsmath}
\usepackage{mathtools}
\usepackage{amsfonts}
\usepackage{siunitx} %
\usepackage[super]{nth}

\usepackage{array}
\usepackage{booktabs}  %
\usepackage{multicol}
\usepackage{multirow}

\usepackage{longtable}
\usepackage[ruled,vlined]{algorithm2e}

\usepackage{newfloat}
\DeclareFloatingEnvironment[
  fileext = lom,
  listname = {List of movies} ,
  name = Movie
]{movie}

\crefname{movie}{Movie}{Movies}
\Crefname{movie}{Movie}{Movies}

\usepackage{siunitx}
\DeclareSIUnit{\molecule}{molecule}
\DeclareSIUnit{\molecules}{molecule}
\DeclareSIUnit\molar{\mole\per\cubic\deci\metre}
\DeclareSIUnit\Molar{\textsc{M}}
\usepackage[shortlabels]{enumitem}
\setlist{nosep} 

\usepackage[dvipsnames, table]{xcolor}

\usepackage[color=green!20]{todonotes}

\usepackage[acronym, nonumberlist, nopostdot, nogroupskip, numberedsection=false]{glossaries}
\setacronymstyle{long-short}
\makeglossaries{}

\newacronym{ddg}{DDG}{Discrete Differential Geometry}
\newacronym{pde}{PDE}{Partial Differential Equation}
\newacronym{fem}{FEM}{Finite Element Method}

\def\Bmu{\mbox{\boldmath$\mu$}}

\def\Bphi{\mbox{\boldmath$\phi$}}

\def\bA{\mbox{\boldmath$ A$}}

\def\bG{\mbox{\boldmath$ G$}}

\def\bL{\mbox{\boldmath$ L$}}

\def\bT{\mbox{\boldmath$ T$}}

\def\bX{\mbox{\boldmath$ X$}}

\def\ba{\mbox{\boldmath$ a$}}

\def\bff{\mbox{\boldmath$ f$}}

\def\br{\mbox{\boldmath$ r$}}

\def\bx{\mbox{\boldmath$ x$}}

\def\vH{\mbox{$\vec{ H}$}}

\def\vK{\mbox{$\vec{ K}$}}

\def\vS{\mbox{$\vec{ S}$}}

\def\va{\mbox{$\vec{ a}$}}

\def\ve{\mbox{$\vec{ e}$}}
\def\vf{\mbox{$\vec{ f}$}}

\def\vn{\mbox{$\vec{ n}$}}

\def\vr{\mbox{$\vec{ r}$}}

\def\vx{\mbox{$\vec{ x}$}}

\def\vba{\mbox{$\vec{\ba}$}}

\def\vbf{\mbox{$\vec{\bff}$}}

\def\vbr{\mbox{$\vec{\br}$}}

\def\Real{\mathbb{R}}
\def\nablar{\nabla_{\vr}}
\def\nablat{\nabla_{\vec{\theta}}}
\def\nablari{\nabla_{\vr_i}}

\def\nablap{\nabla_{\phi}}
\def\nearbyVertices{\sum_{v_{j}\in N (v_i)}}
\def\nearbyEdges{\sum_{e_{ij}\in N (v_i)}}
\def\nearbyHalfedges{\sum_{\underline{e}_{ij}\in N (v_i)}}
\def\nearbyFaces{\sum_{f_{ijk}\in N (v_i)}}
\def\mem3dg{\texttt{Mem3DG}}

\newcommand{\thinsim}{{\raise.17ex\hbox{\(\scriptstyle\mathtt{\sim}\)}}}

\DeclarePairedDelimiter\norm{\lVert}{\rVert}

\begin{document}

\begin{centering}
    \textbf{\Large Mem3DG: Modeling Membrane Mechanochemical Dynamics in 3D using Discrete Differential Geometry}\\[3mm]
    \textbf{C. Zhu$^{1}$, C. T. Lee$^{1*}$, and, P. Rangamani$^{1*}$}\\[1mm]
    \textsuperscript{1}Department of Mechanical and Aerospace Engineering, University of California San Diego, La Jolla CA 92093;\\ 
    $^{*}$To whom correspondence must be addressed: ctlee@ucsd.edu, prangamani@ucsd.edu\\
\end{centering}

\section*{Abstract}
Biomembranes adopt varying morphological configurations that are vital to cellular functions.
Many studies use computational modeling to understand how various mechanochemical factors contribute to membrane shape transformations.
Compared to traditional approximation-based methods (e.g., finite element method), the class of discrete mesh models offers greater flexibility to simulate complex physics and shapes in three dimensions; its formulation produces an efficient algorithm while maintaining expressive coordinate-free geometric descriptions.
However, ambiguities in geometric definitions in the discrete context have led to a lack of consensus on which discrete mesh based model is theoretically and numerically optimal; a bijective relationship between the terms contributing to both the energy and forces from the discrete and smooth geometric theories remains to be established. 
We address this and present an extensible framework, \mem3dg, for modeling 3D mechanochemical dynamics of membranes based on \gls{ddg} on triangulated meshes. 
The formalism of \gls{ddg} resolves the inconsistency and provides a unifying perspective on how to relate the smooth and discrete energy and forces. 
\mem3dg is designed to facilitate models in tandem with and mimicking experimental studies.
It also supports the use of realistic membrane ultrastructure from 3D imaging as an input. 
To demonstrate, \mem3dg is used to model a sequence of examples with increasing mechanochemical complexity: recovering classical shape transformations such as 1) biconcave disk, dumbbell, and unduloid and 2) spherical bud on spherical, flat-patch membrane; investigating how the coupling of membrane mechanics with protein mobility jointly affects phase and shape transformation.
While the first two examples serve as validation, the later examples provide a blueprint for extending \mem3dg to model a system of interest. 
As high-resolution 3D imaging of membrane ultrastructure becomes more readily available, we envision \mem3dg to be applied as an end-to-end tool to simulate realistic cell geometry under user-specified mechanochemical conditions. 
We hope that \mem3dg will be a useful tool to help advance the mission of computational membrane mechanobiology.

\section{Introduction} \label{sec: introduction}
Computational modeling of lipid bilayer mechanics has long been accepted as a way to probe the biophysical aspects of membrane curvature generation. 
The ability of lipid bilayers and cellular membranes to bend in response to various applied forces has been studied extensively from the mathematical modeling perspective.
However, the nonlinear system of equations that result from such modeling often leads to a computational bottleneck to generate predictions from simulations that can be tested against experimentally observed shapes.
In this study, we develop a mesh-based model using discrete differential geometry to reduce this bottleneck.
To justify why our method is necessary and is a computational advance, we first describe the importance of membrane curvature generation in biology, the current state-of-the-art in membrane mechanics modeling, and finally explicitly state the goals of our approach. 

\subsection{Membrane curvature generation in biology}
As one of the most essential and conserved structures of cells, cellular membranes perform many functions.
First, they form compartments to separate chemical environments.
Beyond the passive role of partitioning space, lipids in the membranes interact with proteins and other cellular components influencing cell signaling (e.g., by localizing molecules and acting as an entropic barrier)~\cite{GrovesEtAl2010,ChengEtAl2019}.
Membrane morphology and topology changes are critical for trafficking cargo in and out of cells and are very carefully regulated~\cite{AniteiEtAl2012,BonifacinoEtAl2004,HerrmannEtAl2015,McMahonEtAl2011,ZimmerbergEtAl2006,FarsadEtAl2003}.
Central to these roles is the ability of the membrane to curve and adopt varying morphological configurations from spheres to highly-curved and variegated structures.

Advances in experimental studies of membrane--protein interactions~\cite{AvinoamEtAl2015,terasakiStackedEndoplasmicReticulum2013,ShibataEtAl2010,McMahonEtAl2005,McMahonEtAl2015,KozlovEtAl2014,ScheveEtAl2013,StachowiakEtAl2010,StachowiakEtAl2012,StachowiakEtAl2013,YuanEtAl2021,chabanonSystemsBiologyCellular2017b}, ultrastructural imaging~\cite{KnottEtAl2008,PeddieEtAl2014,TitzeEtAl2016,KnottEtAl2013,MichevaEtAl2007,VillaEtAl2013,Nixon-AbellEtAl2016,HoffmanEtAl2020,XuEtAl2017,FreyEtAl2000}, and image analysis~\cite{WuEtAl2017,lee20203d,mendelsohn2021morphological,SalferEtAl2020,TamadaEtAl2020,AvinoamEtAl2015,terasakiStackedEndoplasmicReticulum2013,ShibataEtAl2010,DaviesEtAl2011,DaviesEtAl2012} have revealed much about the molecular interactions that regulate membrane curvature.
To investigate the mechanics behind these interactions, many theoretical and computational models in terms of membrane energetics and thermodynamics have been developed~\cite{RamakrishnanEtAl2014,helfrich1973elastic,ZimmerbergEtAl2006,steigmann1999fluid,CampeloEtAl2014,naitoNewSolutionsHelfrich1995,desernoFluidLipidMembranes2015,ArgudoEtAl2016,ArgudoEtAl2017,HammEtAl2000,AgrawalEtAl2009,RangamaniEtAl2014,LieseEtAl2020a,LieseEtAl2021,BassereauEtAl2018,BaumgartEtAl2011}.
These models, owing to the ease of \textit{in silico} experimentation, have become an important tool for generating and testing hypotheses~\cite{leeValueModelsMembrane2021a,Carlsson2018}. 
These mechanics models and associated simulations have been used to provide intuition on the mechanical requirements for forming and maintaining complex cellular membrane shapes~\cite{liu2006endocytic,liu2009mechanochemistry,hassinger2017design,alimohamadi2018role,saleemBalanceMembraneElasticity2015,tachikawaGolgiApparatusSelforganizes2017,NatesanEtAl2015,AgrawalEtAl2009a,AgrawalEtAl2010}.

While the utility of this approach has been established and many models have been developed~\cite{RamakrishnanEtAl2014}, many models are limited by critical assumptions or other technical challenges.
For example, the ability to use geometries from membrane ultrastructural imaging experiments as a starting condition would improve model realism~\cite{MaEtAl2021}. 
With respect to computational complexity, the solver should be able to model deformations and topological changes in three dimensions and be compatible with both energy minimization and time-integration for comparing with static and time-series experiments respectively.
This is in contrast to the current assumptions of the existence of an axis of symmetry that is quite commonly made for purposes of ease of simulation~\cite{guckenbergerTheoryAlgorithmsCompute2017}.
An additional feature for these solvers should be that their implementation is modular such that the addition of new physics or increasing model complexity should be straightforward. 
This includes the potential for coupling the membrane model with agent-based and other simulations to propagate other cellular components such as the cytoskeleton~\cite{akamatsu2020principles}.
Thus, new computational tools which are general, easy to use, and without restrictive assumptions are needed to bring modeling closer to experimental observations of membrane shapes in cells.

\subsection{State-of-the-art membrane modeling}
To emphasize the motivations behind our choice of extending and developing a new mesh-based membrane model, we provide a brief summary of the legacy literature in modeling membrane mechanics. 
The most common theoretical model of membrane bending is the Helfrich-Canham-Evans Hamiltonian,\footnote{The Helfrich energy is related to the Willmore energy in the mathematics literature~\cite{willmore1996riemannian}} which describes the lipid bilayer as a two-dimensional fluid-like structure that exhibits solid-like elasticity in the out-of-plane direction~\cite{helfrich1973elastic,steigmann1999fluid,evans2018mechanics,canham1970minimum,jenkins1977equations}.
It is a continuum model which describes the bending energy of the membrane as a function of its mean and Gaussian curvatures.
The assumptions for the continuum are satisfied as long as the deformations are much larger in length scale compared to the individual lipid components.

Given the necessary material properties and boundary conditions, by minimizing the Helfrich energy, we can obtain the equilibrium shape of the membrane~\cite{jenkins1977equations,jenkins1977static,seifert1997configurations,helfrich1973elastic}.
While straightforward in concept, energy minimization requires the determination of the forces on the membrane which is a challenging task~\cite{guckenbergerTheoryAlgorithmsCompute2017}.
The forces on the membrane are given by the variation of the energy with respect to the embedded coordinate (i.e., shape) of the membrane\footnote{We call this variation the shape derivative which is distinct from the chemical derivative that will be introduced later in the context of mechanochemical coupling}.
Taking the shape derivatives of the Helfrich energy produces the ``shape equation'', so termed because solutions of this partial differential equation, with the prescribed boundary conditions, produce configurations at equilibrium (i.e., force-balance).

Solving the shape equation is non-trivial since it is a partial differential equation with fourth-order nonlinear terms.
As a result, analytical solutions of the shape equation are known only for a few cases constrained to specific geometries and boundary conditions~\cite{naitoNewSolutionsHelfrich1995}.
For most systems, we must resort to the use of numerical methods.
The simplest numerical schemes can be formulated by making restrictive assumptions such as considering only small deformations from a plane (e.g., Monge parametrization) or assuming that there exists an axis of symmetry such that the resulting boundary value system can be integrated~\cite{RamakrishnanEtAl2014}.
While these methods are suitable for idealized shapes, these assumptions are not consistent with the membrane shapes found in biology are and thus not general enough for advancing the field.

\begin{table}[htbp]
    \caption{
        Comparison of common mathematical frameworks for modeling membrane mechanics with specifications to advance the mission of computational membrane mechanobiology.
        A general framework will permit the easy transfer of inputs and results between model and experiments.
        Models which can be coupled with other modeling schemes representing other cellular components can help address the complexity of cell biology.
        Discrete mesh models have many desirable traits, with respect to these specifications, at the cost of forgoing rigorous error analysis.
    }\label{table: state of the arts} 
    \begin{tabular*}{\textwidth}
        {@{}m{0.55\textwidth}*{3}{>{\centering\arraybackslash}m{0.12\textwidth}}@{}}
        \toprule
                                                                      & \textbf{Phase field/level set} & \textbf{FEM}                     & \textbf{Discrete Mesh/\mem3dg} \\ \midrule
        \textbf{General 3D}                                           & \checkmark                     & \checkmark                       & \checkmark                     \\ \midrule 
        \textbf{Statics + dynamics}                                   & \checkmark                     & \checkmark                       & \checkmark                     \\ \midrule 
        \textbf{Membrane heterogeneity}                               & \checkmark                     & \checkmark                       & \checkmark                     \\ \midrule 
        \textbf{Incorporation of agents/particles}                    &                                &                                  & \checkmark                     \\ \midrule 
        \textbf{Incorporation of stochastic dynamics (e.g., DPD or MC)} &                                &                                  & \checkmark                     \\ \midrule 
        \textbf{Explicit surface parametrization}                     &                                & \checkmark                       & \checkmark                     \\ \midrule 
        \textbf{Coordinate-free evaluation}                               &                                &                                  & \checkmark                     \\ \midrule 
        \textbf{Ability to support topological changes}               & \checkmark                     &                                  & requires mesh surgery           \\ \midrule 
        \textbf{Error analysis}                                       & \checkmark                     & \checkmark                       &                                \\ \bottomrule
    \end{tabular*}
\end{table}

Solvers of membrane shape in 3D have also been developed and can be categorized into three groups: 1) phase field or level set methods~\cite{DuEtAl2004,du2006convergence,salac2011level}, 2) \gls{fem}~\cite{fengFiniteElementModeling2006, ma2008viscous, elliottModelingComputationTwo2010, rangarajanFiniteElementMethod2015, sauerStabilizedFiniteElement2017,vasanMechanicalModelReveals2020,torres-sanchezModellingFluidDeformable2019,auddyaBiomembranesUndergoComplex2021}, and 3) discrete surface mesh models~\cite{gompper1996random, julicher1996morphology,kantor1987phase,bianBendingModelsLipid2020,brakkeSurfaceEvolver1992,jieNumericalObservationNonaxisymmetric1998, krollConformationFluidMembranes,atilganShapeTransitionsLipid2007a,bahramiFormationStabilityLipid2017,noguchiShapeTransitionsFluid2005a, tachikawaGolgiApparatusSelforganizes2017,tsaiRoleCombinedCell2020}.
These methods and others, reviewed in detail by \textcite{guckenbergerBendingAlgorithmsSoft2016}, differ in the strategy used to discretize the membrane domain and compute the relevant derivatives.
We compare the aforementioned general, 3D models to our established model criteria in \Cref{table: state of the arts} and elaborate below.

Phase field and level set methods solve the shape equation by propagating a density field on an ambient volumetric mesh.
The membrane shape is implicit in these models and can be found by drawing an isosurface or level set of the model.
While this is ideal for modeling membrane topological changes, the implicit representation of the membrane adds complexity for interfacing with data generated using modern methods of visualizing membrane ultrastructure.
The meshes output from ultrastructural studies must be converted into a density or phase field prior to input to the model.
While this conversion is possible, representing the dynamic and variegated shapes of cellular membranes would require a dense volume mesh, which reduces computational tractability.
The implicit surface representation also complicates the addition of new in-plane physics for end-users.

\gls{fem} and discrete mesh models use an explicit surface parametrization (i.e., a mesh).
Thus the meshes output from ultrastructural imaging datasets can be used in these frameworks with minor modifications~\cite{lee20203d,Shewchuk2002}.
Comparing \gls{fem} methods with our specifications we identify a few key challenges.
First, the numerical evaluation of smooth geometric measurements on arbitrary manifolds in an \gls{fem} framework requires non-intuitive tensor algebra to translate the shape equation in coordinate where it is ready to be solved.
After this formulation, solving the shape equation can require the use of high order function basis such as the \(C^1\) conforming \gls{fem} based on subdivision scheme~\cite{fengFiniteElementModeling2006,ma2008viscous} or isogeometric analysis (IGA)~\cite{sauerStabilizedFiniteElement2017,vasanMechanicalModelReveals2020,auddyaBiomembranesUndergoComplex2021,rangarajanFiniteElementMethod2015}, which adds code complexity and run-time cost.
Extending an \gls{fem} framework to incorporate new physics, topological changes, or interfaces with other models requires advanced mathematical and coding skills.
This can restrict the usage to the computational math community and prevent broad usage by the biophysics community.

Finally, evaluating discrete mesh-based methods, which define the system energy and/or forces using geometric primitives from a mesh, we find that they satisfy many of the requirements in \Cref{table: state of the arts}.
Due to the ease of use and implementation, discrete mesh models have gained in popularity and many different schemes can be found in the literature~\cite{gompper1996random, julicher1996morphology,kantor1987phase,bianBendingModelsLipid2020,brakkeSurfaceEvolver1992,jieNumericalObservationNonaxisymmetric1998,krollConformationFluidMembranes,atilganShapeTransitionsLipid2007a,bahramiFormationStabilityLipid2017,noguchiShapeTransitionsFluid2005a, tachikawaGolgiApparatusSelforganizes2017,tsaiRoleCombinedCell2020,pezeshkian2019multi,sadeghiParticlebasedMembraneModel2018}.
These schemes differ in their approach to defining and computing geometric measurements necessary for defining the energy and forces on a discrete object.
Discrete geometries have discontinuities and limited information that leads to degenerate definitions for geometric values. 
For example, there is no canonical definition for the normal of a vertex of a mesh as opposed to the normal of a smooth geometry~\cite{chernDDG,craneDISCRETEDIFFERENTIALGEOMETRY,bianBendingModelsLipid2020}.
One challenge for selecting the suitable formulation to use is the lack of approximation error metric for which the discrete definition best matches the smooth theory.
Another confounding factor is the step at which the problem is discretized.
Some implementations discretize the energy of the system by constructing standalone discrete energy, which captures the behavior of the Helfrich energy~\cite{guckenbergerTheoryAlgorithmsCompute2017}.
From this discrete energy, they take the shape derivatives to obtain an expression for the discrete force.
Without careful consideration, the discrete forces derived in this manner are unstructured and there is little resemblance to expressions of force from smooth theory.
A second option is to discretize the smooth force expression directly\cite{guckenbergerBendingAlgorithmsSoft2016,guckenbergerTheoryAlgorithmsCompute2017}.
While this preserves the geometric connection for the forces, there is no longer well-defined discrete energy.
Several discrete mesh methods were benchmarked by \textcite{bianBendingModelsLipid2020} and \textcite{guckenbergerBendingAlgorithmsSoft2016} who found differences in the accuracy, robustness, and ease of implementation~\cite{bianBendingModelsLipid2020,guckenbergerBendingAlgorithmsSoft2016}.

\subsection{Goals of the current work}
In this work, we outline a discrete mesh framework for modeling membrane mechanics with the following goals in mind: (a) we do not make \textit{a priori} assumptions about axes of symmetry or restrict the coordinates in any way; (b) we resolve the ambiguity in the definition of geometric measurements on the mesh and permit direct comparison for both the energy and force expressions in smooth and discrete contexts; and (c) this framework allows for use of meshes generated from ultrastructural imaging. 
We begin by defining a discrete energy that is analogous to the Helfrich energy.
Then using concepts from \gls{ddg}, we derive discrete shape derivatives and group terms to produce a discrete shape equation.
We will show that our discrete shape equation has a clear correspondence between the terms of the smooth shape equations~\cite{jenkins1977equations,jenkins1977static,hassinger2017design,willmore1996riemannian}.
Beyond establishing this important connection, we will show that the elegant analytical expressions for discrete variational terms from the \gls{ddg} also yield improved geometric intuition and numerical accuracy~\cite{chernDDG,craneDISCRETEDIFFERENTIALGEOMETRY}.

Benchmarking of our expressions was performed with our accompanying software implementation called Membrane Dynamics in 3D using Discrete Differential Geometry (\mem3dg). 
\mem3dg is written in C++, released under the Mozilla Public License version 2, and comes with accompanying documentation and tutorials which can be accessed on GitHub (\url{https://github.com/RangamaniLabUCSD/Mem3DG}).
Beyond the computation of discrete energies and forces on a mesh of interest, we also include functionality for performing energy minimization and time integration.
Using \mem3dg, we validate the exactness of the analytical expressions of force terms by numerically examining the convergence of the force terms as a function of system energy perturbation.
To illustrate compliance with our tool specifications, we apply \mem3dg to a sequence of examples with increasing complexity.
Finally, we outline the steps to incorporate additional physics such as membrane-protein interactions and surface diffusion into \mem3dg.

\section{Methods}\label{sec: method}

The lipid bilayer is modeled as a thin elastic, incompressible shell using the 
Helfrich-Canham-Evans Hamiltonian or spontaneous curvature model~\cite{helfrich1973elastic,canham1970minimum,Evans1974}.
The bending energy, $E_b$, of a smooth surface or 2-manifold, $\mathcal{M}$, can be expressed in terms of the mean $H$, Gaussian $K$, and spontaneous curvature $\bar{H}$ with material parameters $\kappa$ the bending and $\kappa_G$ the saddle-splay moduli.
Additional energy terms $E_s$ and $E_p$ account for the tension--area ($\lambda$--$A$) and pressure--volume ($\Delta P$--$V$) relationships; 
The total energy of the bilayer is therefore
\begin{equation}\label{eqn: smooth free energy}
    E = \underbrace{\int_{\mathcal{M}}\left[\kappa(H-\bar{H})^{2}+\kappa_{G} K\right] d A}_{E_b}  +  \underbrace{\int_{\bar{A}}^{A} \lambda d \tilde{A}}_{E_s} - \underbrace{\int_{\bar{V}}^{V} \Delta P d \tilde{V}}_{E_p}.
\end{equation}
The preferred surface area and volume, $\bar{A}$ and $\bar{V}$, combined with the spontaneous curvature, $\bar{H}$, characterize the zero-energy state for the system energy. 
In a nutshell, given the material properties, the system energy is fully determined by its geometric measurements such as volume, area, and curvatures. 

Machinery to express these measurements have been a topic of extensive study in classical differential geometry~\cite{do2016differential,Lee2012a}.
However, finding the minima of the governing energy, solving the stationary solution to the geometric \gls{pde}, can be mathematically and numerically difficult. 
While differential geometry provides succinct expressions to describe the measurements in a coordinate-free fashion, computational methods often require the introduction of a coordinate basis and subsequent manipulation of expressions using tensor algebra, which can obscure the underlying geometric intuition. 

As an alternative, forgoing the need for a smooth geometry, one can treat a discrete geometry (such as a geometric mesh) as the input.
This perspective where the discrete geometry is the actual geometry is that of \gls{ddg}\cite{bobenkoDiscreteDifferentialGeometry2008}.
By eliminating the burdens of treating the input mesh as an approximation of a smooth object, \gls{ddg} capitalizes upon the piecewise nature of meshes to produce efficient and parallelizable finite difference-like formulae which are amenable to algorithmic implementation while maintaining clear geometric meaning.
In the following sections, we use concepts from \gls{ddg} to formulate a discrete analog to the smooth membrane shape problem.
Following the derivation of the discrete theory, we describe the development of an accompanying software implementation called \mem3dg.

\begin{table}[htbp]
    \caption{Glossary of commonly used symbols and conventions}
    \label{table: symbols}

    \begin{minipage}[t]{0.5\linewidth}
        \textbf{A. Geometric primitives:}\\[2mm]
        \begin{tabular}{@{}p{0.2\linewidth}p{0.7\linewidth}@{}}
            $\mathcal{M}$ & smooth or discrete 2-manifold\\
            $\vr \in \Real^3$ & embedded coordinate of $\mathcal{M}$\\
            $l$ & edge length \\
            $\angle $ & corner angle \\
            $\varphi$ & dihedral angle  \\
            $A$ & area of mesh cell, \\
            & e.g., face $A_{ijk}$, edge $A_{ij}$ and vertex $A_{i}$ \\
            $\vn$ & surface normal \\
        \end{tabular}\\[2mm]
        \textbf{B. Surface Integral:}\\[2mm]
        \begin{tabular}{@{}p{0.2\linewidth}p{0.7\linewidth}@{}} 
            $\int a $ & integrated quantity over mesh cell e.g., $A_i a_i$ or $A_{ijk} a_{ijk}$\\
            $\sum_{v_i}$ & sum over all vertices $v_i$ of the mesh\\
            $\sum_{e_{ij}}$ & sum over all edges $e_{ij}$ of the mesh\\
            $\sum_{f_{ijk}}$ & sum over all faces $f_{ijk}$ of the mesh\\
            $\sum_{v_{j}\in N(a)}$ & sum over the vertex $v_{j}$ in the neighborhood of $a$\\
            $\sum_{e_{ij}\in N(a)}$ & sum over the edges $e_{ij}$ in the neighborhood of $a$\\
            $\sum_{f_{ijk}\in N(a)}$ & sum over the face $f_{ijk}$ in the neighborhood of $a$\\
        \end{tabular}\\[2mm]
        \textbf{C. Tensors:}\\[2mm]
        \begin{tabular}{@{}p{0.2\linewidth}p{0.7\linewidth}@{}}
            $x \in \Real$ & scalar quantity \\
            $x^{\text{type}}_{\text{index}}$ & sub- and super-script convention e.g., $\int \vf^b_i$ is the bending force for vertex $i$ \\
            $\vx \in \Real^3$ & vector quantity\\
            $\bx = \{x_i\}$ & $(n \times 1)$ indexed scalar quantity\\
            $\Vec{\bx}= \{\bx_i\}$ & $(n \times 3)$ indexed vector quantity \\
            $\tilde{\bX}$ & matrix or tensor quantity\\
         \end{tabular}
    \end{minipage}\hfill
    \begin{minipage}[t]{0.5\linewidth}
        \textbf{D. Derivatives:}\\[2mm]
        \begin{tabular}{@{}p{0.175\linewidth}p{0.75\linewidth}@{}}
            $\nablar$ & shape derivative\\
            $\nablap$ & chemical derivative \\
            $\nablat$ & surface gradient \\
            $\dot{a}$ & time derivative \\
            $\Delta_s$ & Laplace-Beltrami operator \\
        \end{tabular}\\[2mm]
        \textbf{E. Physical Variables:}\\[2mm]
        \begin{tabular}{@{}p{0.175\linewidth}p{0.75\linewidth}@{}}
            $E$ & energy \\
            $f$ &force density \\
            $\mu$ & chemical potential \\
            $H$ & mean curvature \\
            $K$ & Gaussian curvature \\
            $A$ & surface area\\
            $V$ & enclosed volume \\
            $\bar{\cdot}$ & preferred state e.g., $\bar{H}$ is the spontaneous curvature\\
            $\phi \in [0, 1]$ & protein density parameter \\
            $\lambda$ & membrane tension \\
            $\Delta P$& osmotic pressure across the membrane \\
            $\kappa$ & bending rigidity \\
            $\kappa_G$ & Gaussian modulus \\
            $K_A$ & stretching modulus \\
            $K_V$ & osmotic strength constant \\
            $\bar{c}$ & molar ambient concentration \\
            $n$ & molar quantity of enclosed solute \\
            $\eta$ & Dirichlet energy constant \\
            $\varepsilon$ & adsorption energy constant \\
            $\xi$ & membrane drag constant \\
            $B$ & protein mobility constant \\
        \end{tabular}
    \end{minipage}
\end{table}

\subsection{Notation and preliminaries}
\label{sec: notation}

We assume the following notation conventions and provide a table of important symbols (\cref{table: symbols}). 
To aid the reader on how the elements of the mesh are used in the derivation, several fundamental geometric primitives (i.e., values on a mesh which are easily measurable; listed in \cref{table: symbols}A) are illustrated in \cref{fig: energy and force}A-C.

We note that in discrete contexts the notation, $\int a$, should be considered the discrete (integrated) counterpart of a pointwise measurement $a$ in a smooth setting.
The rationale and significance behind the usage of an integrated quantity in discrete contexts are elaborated in \Cref{sec:integrated_measures} and the \gls{ddg} literature~\cite{chernDDG,craneDISCRETEDIFFERENTIALGEOMETRY}.
Using this notation, discrete surface integrals are expressed as sums of integrated values over the discrete mesh components listed in \cref{table: symbols}B (e.g., \(\sum_{v_i}\int a_i\) is the discrete analog to \(\int_{\mathcal{M}}a\)).
It is possible to interchange between integrated, $\int a_i$, and pointwise, $a_i$, quantities by using the dual area ($A_i$), 
\begin{equation}
    \label{eqn: relating pointwise and integrated measures}
    a_i = \int a_i/A_i.
\end{equation}
For simplicity, we will not use separate notations for operators applying in smooth and discrete settings.
The context can be inferred from the objects  to which the operators are applied.
Where it serves to improve our geometric or other intuition, smooth objects will be presented alongside discrete objects for comparison.

\subsection{Obtaining a discrete energy defined by mesh primitives}
\label{sec: discrete energy defined by mesh primitives}

Following the perspective of \gls{ddg}, we restrict our input to the family of triangulated manifold meshes, $\mathcal{M}$ (i.e., discrete 2-manifolds embedded in $\Real^3$)\footnote{We will use $\mathcal{M}$ for both the smooth and discrete surfaces.}.

Paralleling the smooth Helfrich Hamiltonian (\cref{eqn: smooth free energy}), a functional of geometric measurements of a surface, the discrete Helfrich Hamiltonian is composed of discrete analog of those measurements,
\begin{equation}\label{eqn: discrete free energy}
    E(\vbr) = \underbrace{ \sum_{v_i} \left[  \kappa_i \int (H_i - \bar{H}_i)^2 + \kappa_G \int K_i \right] }_{E_b} + \underbrace{\int_{\bar{A}}^A \lambda(\tilde{A}; \vbr) ~ d \tilde{A}}_{E_s} - \underbrace{\int_{\bar{V}}^{V}\Delta P(\tilde{V}; \vbr) ~ d \tilde{V}}_{E_p}.
\end{equation}
In comparison with \cref{eqn: smooth free energy}, $H_i$ and $K_i$ are pointwise mean and Gaussian curvature measurements on vertices, $\int (H_i - \bar{H}_i)^2$ is the integrated Willmore measure, and the smooth surface integral is replaced by its discrete analog (i.e. finite summation), $\sum_{v_i}$ (\cref{table: symbols}B).

The geometric properties of a given membrane configuration can be connected to the system's energy through constitutive relations.
In this work, we assume that the surface tension follows a linear stress-strain model\cite{Phillips2009}, 
\begin{equation}\label{eqn: surface tension}
    \lambda (A;\vbr) = K_A \frac{A(\vbr) - \Bar{A}}{\Bar{A}},
\end{equation}
where $\bar{A}$ is the preferred surface area of the membrane, and $K_A$ is the stretching modulus of the membrane. 
The osmotic pressure can be defined based on the van't Hoff formula as
\begin{equation}\label{eqn: osmotic pressure}
    \Delta P (V;\vbr) =  P_{\text{in}} - P_{\text{out}} = iRT \left(\frac{n}{V} - \bar{c}\right),   
\end{equation}
where $i$, $R$, $T$, $\bar{c}$ and $n$ are the van't Hoff index, ideal gas constant, temperature, ambient molar concentration, and molar amount of the enclosed solute.
Substituting these constitutive relations (\cref{eqn: surface tension,eqn: osmotic pressure}) into the energy (\cref{eqn: discrete free energy}), we get explicit expressions for $E_s$ and $E_p$, 
\begin{equation}\label{eqn:discrete_energy_with_relations}
    E(\vbr) =  E_b(\vbr) + \underbrace{\frac{1}{2} K_A \frac{[A(\vbr) - \Bar{A}]^2}{\Bar{A}}}_{E_s} +
    \underbrace{i R T n \left[ r_c - \ln{r_c} - 1\right]}_{E_p},
\end{equation}
where $r_c = \bar{c}/(n/V)$ is the ratio of the concentrations of the ambient and enclosed solutions.
Note that the preferred volume, $\bar{V}$, which is needed to evaluate the integral in \cref{eqn: discrete free energy}, is related to to the parameters in \cref{eqn: osmotic pressure} by $\bar{V} = n/\bar{c}$.
If the system is around the isosmotic condition (e.g., $V \rightarrow \bar{V}$), the leading order of the energy is given as,
\begin{equation}\label{eqn: osmotic work-phenomenological}
    E_p \approx \frac{1}{2} K_V \frac{(V - \bar{V})^2}{\bar{V}^2},
\end{equation}
where $K_V \equiv iRTn$ groups the phenomenological parameters.

\begin{figure}[htbp]
    \centering
    \includegraphics[width=\linewidth]{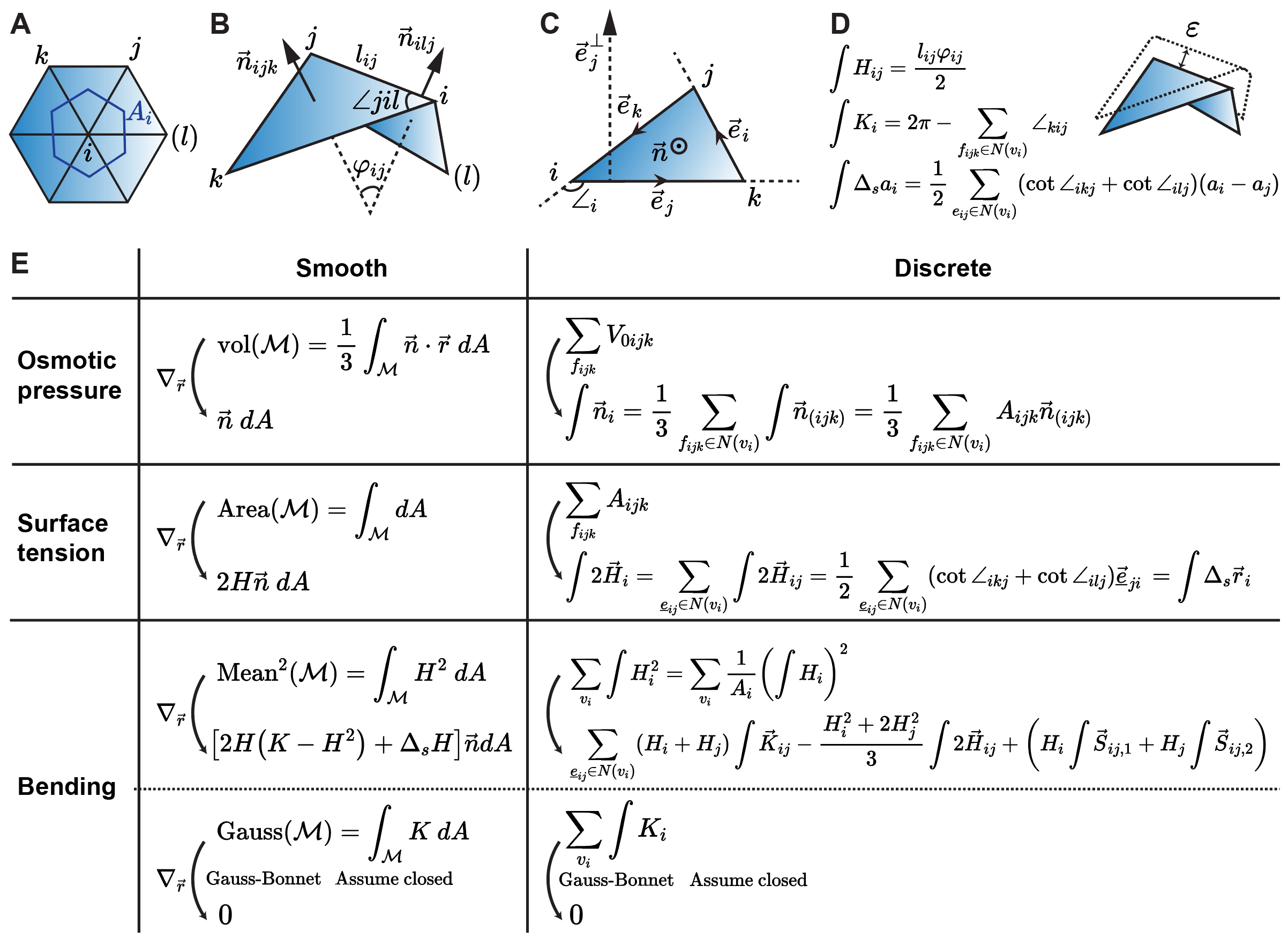}
    \caption{Overview of the \gls{ddg} framework. A, B, C) Illustrations of geometric primitives in the neighborhood of A) Fan around a vertex, B) Diamond around an edge, and C) Triangle on a face. D) Discrete definition of scalar edge mean curvature, $\int H_{ij}$, scalar vertex Gaussian curvature, $\int K_i$, and Laplace-Beltrami operator, $\int \Delta_s (\cdot)$. E) Comparative derivation of Helfrich shape equation in both smooth and discrete formulation.}
    \label{fig: energy and force}
\end{figure}

To compute the energy of a system, we must obtain values for several geometric measurements which appear in the discrete energy function (i.e., $H$, $K$, $A$, $V$, etc.).
For measurements such as the volume and area, there are clear approaches for their evaluation on a triangulated mesh: summing the areas of the triangular faces and summing over the signed volume of tetrahedra (\cref{fig: energy and force}E -- osmotic pressure and surface tension).
For other measurements such as the discrete mean and Gaussian curvatures, additional care must be taken.
While in smooth contexts these curvatures have unique definitions, in discrete contexts there are multiple approaches for their calculation.
For example, the mean curvature can be computed via the application of the cotangent Laplacian, the kernel of the heat equation, or fitting polynomials to a local patch~\cite{guckenbergerTheoryAlgorithmsCompute2017}.
As mentioned earlier, there are challenges with these approaches which can limit their numerical accuracy and obscure the connection to smooth theory.
Here using discrete exterior calculus and identification of geometric invariants, we produce theoretically and numerically consistent discrete expressions.

Similar to the polygonal curve introduced in \cref{sec:integrated_measures}, a triangulated mesh has zero curvature on facets and ill-defined curvature on edges and vertices.
Using the Steiner view, the sharp corners formed by vertices and edges are made smooth with portions of spherical and cylindrical shells, which have well-defined mean curvature (\cref{fig: energy and force}D).
Taking the limit as the radii of the cylinders and spheres decrease, the leading order contribution of total mean curvature is given by the Steiner formula on an edge,
\begin{equation}\label{eqn: scalar mean curvature}
    \int H_{ij}  = \frac{l_{ij}\varphi_{ij}}{2},
\end{equation}
referred to as the edge mean curvature, where $l_{ij}$ is the length of edge $e_{ij}$, and $\varphi_{ij}$ is the dihedral angle on $e_{ij}$ (i.e., the angle formed by the face normals of the neighboring triangles incident to $e_{ij}$) (illustrated in \cref{fig: energy and force}B)\cite{chernDDG, craneDISCRETEDIFFERENTIALGEOMETRY}.
While not necessary, a triangulated mesh is often realized in $\Real^3$ via vertex positions; thus it is conventional to prescribe data on vertices instead of edges. 
Summation of edgewise quantities over the ``fan" neighborhood (\cref{fig: energy and force}A) provides the recipe of converting an edgewise to a vertexwise quantity,
\begin{equation}\label{eqn: edge value to vertex value}
    (\cdot)_i = \frac{1}{2} \nearbyEdges (\cdot)_{ij},
\end{equation}
where the prefactor, $1/2$, accounts for fact that each edge is shared by two vertices.

While we have an integrated mean curvature, the discrete Helfrich Hamiltonian contains a pointwise mean curvature squared term.
To define a pointwise mean curvature, the size of the domain \textit{occupied} by the integrated mean curvature needs to be specified (c.f.,  \cref{sec:integrated_measures} for rationale). 
The area, $A_{i}$, referred to as the dual area of the vertex $v_i$, can be defined as one-third of the areal sum of the incident triangles (``fan'' illustrated in \cref{fig: energy and force}A)\cite{chernDDG, craneDISCRETEDIFFERENTIALGEOMETRY}.
Applying \cref{eqn: relating pointwise and integrated measures,eqn: edge value to vertex value} to \cref{eqn: scalar mean curvature}, the pointwise mean curvature is thus,
\begin{equation}\label{eqn: pointwise mean curvature}
    H_{i} = \frac{\int H_{i}}{A_{i}} = \nearbyEdges \frac{l_{ij}\varphi_{ij}}{4 A_i}.
\end{equation}
Substituting \cref{eqn: pointwise mean curvature} into the integrated Willmore measure term of \cref{eqn: discrete free energy}, the integrated Willmore measure can be expressed as a function of the integrated mean and spontaneous curvature,
\begin{equation}
    \int (H_i - \bar{H}_i)^2 =  \frac{1}{A_i} \left(\int H_i - \int \bar{H}_i
    \right)^2.
\end{equation}
Discrete Gaussian curvature is given by the angle defect formula, 
\begin{equation}\label{eqn:integrated_gaussian_curvature}
 \int K_i = 2 \pi - \nearbyFaces \angle_{kij}, 
 \end{equation}
which is a well-known quantity that preserves many properties parallel to the smooth theory (e.g., Gauss-Bonnet, turning number, among other invariants).
One way to derive the angle defect formula is to compute the area of a spherical $n$-gon contained by a local Gauss map of the neighboring $n$ faces around a vertex ~\cite{chernDDG, craneDISCRETEDIFFERENTIALGEOMETRY}. 
The discrete Gauss-Bonnet theorem states that
\begin{equation}
    \sum_{v_i} \int K_i = 2\pi \chi(\mathcal{M}) - \sum_{e_{ij} \in \partial \mathcal{M}} \int \kappa^g_{ij},
\end{equation}
where $\chi(\mathcal{M}) = |V| - |E| + |F|$ is the Euler characteristic of $\mathcal{M}$, a topological invariant where $|V|$, $|E|$ and $|F|$ represent the number of vertices, edges and faces of the mesh respectively.
The discrete geodesic curvature, $\int \kappa^g$, is the discrete curvature of the boundary curve, $\partial \mathcal{M}$, as introduced and defined in \cref{sec:integrated_measures}.
When the surface $\mathcal{M}$ is closed and does not undergo topological changes, the boundary term drops out and the total Gaussian curvature is a constant multiple of $2 \pi$.
When surface $\mathcal{M}$ has boundaries, the geodesic curvature is integrated over one or more closed boundary loops, $\partial \mathcal{M}$.
This integral, referred to as the turning number of a closed polygon, is invariant under regular homotopy (i.e., continuous deformation during which the curve stays regular)~\cite{chernDDG}, which is admitted by the deformation of all membrane boundaries considered in this study.
In summary, within the scope of the current work, all energetic contributions from Gaussian curvature terms, including the boundary elements, are constant or zero and can thus be neglected.
Nevertheless, in future studies where Gaussian curvature term cannot be neglected (e.g, topological changes, non-uniform saddle-splay modulus), \cref{eqn:integrated_gaussian_curvature} provides the geometric definition to obtain the discrete energy.

A numerical comparison of the discrete scalar measurements with their smooth counterparts is shown in \cref{SI_fig: ptwise magnitude}. 
We note that for all geometric measurements (i.e., volume, area, and curvatures), unlike in smooth differential geometry where their numerical evaluation requires the introduction of coordinates, \gls{ddg} measurements are functions of mesh primitives. 
By substituting these discrete geometric measurements from \gls{ddg} into \cref{eqn:discrete_energy_with_relations} and \cref{eqn: discrete free energy}, we get a numerical recipe for computing the total system energy.

\subsection{Force from discrete shape derivative of energy}
\label{sec: force from discrete shape derivative of energy}

We can obtain the force by taking the negative shape derivative of the energy.
In continuous settings, the differentiation is an infinite-dimensional problem that requires the use of the calculus of variations and differential geometry to find analytical expressions~\cite{jenkins1977equations,jenkins1977static, helfrich1973elastic, desernoFluidLipidMembranes2015} (\cref{fig: energy and force}E -- smooth).
Deriving the forces from the discrete energy (\cref{eqn: discrete free energy}) is a much simpler task.

Discrete forces can be obtained by taking partial derivatives of mesh primitives with respect to the vertex embedded coordinates, $\vr$ (\cref{fig: energy and force}E -- discrete).
Regarding notation, despite the computational differences, the differential operations in both the discrete and smooth contexts are called (discrete) shape derivatives and denoted as $\nablar(\cdot)$ due to the common geometric meaning.
We note that the computation of discrete shape derivatives for membrane modeling has been described previously in the literature~\cite{bianBendingModelsLipid2020, julicher1996morphology}.
Also that there are many overlapping definitions for discrete curvature, energy, and variations thereof in the graphics literature~\cite{meyerDiscreteDifferentialGeometryOperators2003, wardetzkyDiscreteLaplaceOperators2008, wardetzkyConvergenceCotangentFormula2008}.
Our work extends upon the prior art which evaluates derivatives algebraically, by introducing simplifications based upon the grouping of terms and identification of geometric objects.
These simplifications have important implications for improving the geometric understanding as well as the run-time and numerical performance of an implementation.

At the high level, our goal is to express the forces on each vertex, given a set of physics, using geometric primitives and properties defined on specific mesh elements.
By grouping terms, we find that the vertex-wise forces arising from the different physics can be expressed as weights which are functions of the input parameters and system configuration, multiplied by basic geometric vectors.
We will show that these terms have an exact correspondence to terms in the smooth shape equation (\cref{fig: energy and force}E).
We remark that, in some sense, the force expressions are reminiscent of finite difference equations which approximate differentials as a linear combination of values at discrete points.
This may have implications for the suitability of modeling smooth biological surfaces with discrete meshes.

\subsubsection{Force from osmotic pressure}
For the smooth geometry, the shape derivative of the enclosed volume yields the outward pointing surface normal with its size equal to the local area element \cite{desernoNotesDifferentialGeometry}.
For a discrete mesh, the shape derivative of the volume is given by the face normal on \emph{triangular faces} with its local area element equaling to the \emph{face area}, which is referred to as the \emph{integrated} face normal, $ \int \vn_{(ijk)}$ (\cref{fig: energy and force}E -- osmotic pressure)~\cite{bianBendingModelsLipid2020,craneDISCRETEDIFFERENTIALGEOMETRY,chernDDG}, where $(ijk)$ denotes the symmetry under index permutation (e.g., $a_{i(jk)}$ means $a_{ijk} = a_{ikj}$).
Similar to edge values, the force normal can be converted to vertex normal,
\begin{equation}
     \int \vn_i = \nablari V =\frac{1}{3} \nearbyFaces \int \vn_{(ijk)} = \frac{1}{3} \nearbyFaces A_{ijk}\vn_{(ijk)},
\end{equation}
where analogous to \cref{eqn: edge value to vertex value}, the prefactor $1/3$ accounts for fact that each face is shared by three vertices.
The discrete vertex forces from the derivative of the pressure-volume work, $\int \vf^{\,p}_i$, is then given by scaling it with the uniform osmotic pressure, 
\begin{align}\label{eqn: discrete osmotic force}
     \int \vf^{\,p}_i = \Delta P \int \vn_i.
\end{align}

\subsubsection{Forces from surface tension}
Next considering the shape derivative of the surface energy, $E_s$, in smooth contexts, the derivative of the total surface area also points at the surface normal, with its magnitude measuring the size ($dA$) and the curvature ($2H$) of the local patch (\cref{fig: energy and force}E -- surface tension) \cite{desernoNotesDifferentialGeometry}.  
In discrete case, we can directly compute the derivative of total area on each vertex by summing the area gradient of incident triangles with respect to the vertex position; The sum is therefore referred as (twice of) the integrated mean curvature vector on vertices,
\begin{align}
        \int 2 \vH_i = \nablari A  = \nearbyFaces \nablari A_{ijk} = \nearbyFaces  \int 2 \vH_{i(jk)},
\end{align}
where we define $\int 2 \vH_{i(jk)} \equiv \nablari A_{ijk}$, and $\int \vH_{i(jk)}$ is the mean curvature vector on a triangle face corner. 
The capillary forces from surface tension, $\int \vf^{\,s}_i$, are given by scaling the integrated mean curvature vector by the surface tension,
\begin{align}\label{eqn: discrete capillary force}
     \int \vf^{\,s}_i = - \lambda \int 2 \vH_i.
\end{align}

Evaluating the algebraic sum of area gradients reveals the ``cotangent formula'' applied to the vertex positions (\cref{fig: energy and force}E -- surface tension). 
From independent derivations with unrelated frameworks (e.g., discrete exterior calculus and \gls{fem}), discretizing the smooth Laplace-Beltrami operator on a triangulated mesh produces the cotangent formula which is called the discrete Laplace-Beltrami operator, $\int \Delta_s$\cite{craneDISCRETEDIFFERENTIALGEOMETRY,meyerDiscreteDifferentialGeometryOperators2003,chernDDG}.
Inspecting the expressions in \cref{fig: energy and force}E -- surface tension, we see that our discrete expression parallels smooth theory, where the mean curvature is related to the coordinates through the application of the smooth Laplace-Beltrami operator,
\begin{equation}
    \nearbyEdges \int 2\vH_{ij} = \int \Delta_s \vr_i \quad \leftrightarrow \quad \Delta_s \vr = 2H \vn.
\end{equation}

\subsubsection{Forces from bending}\label{sec: forces from bending}

To evaluate the shape derivative of the discrete bending energy we consider the terms from the Gaussian and mean curvature separately.
Based on the discrete Gauss-Bonnet theorem (\cref{sec: discrete energy defined by mesh primitives}), the total Gaussian curvature only varies if the surface undergoes a topological change.
Since we do not consider non-uniform saddle-splay modulus and topological changes in the examples in this work, this term does not contribute to the force.
The shape derivative of the bending energy contributed by the integrated Willmore measure requires more algebra and the introduction of halfedges, $\underline{e}_{ij}$ (c.f., \cref{sec: halfedge}).
Here we will focus on the key results and refer the reader to the full derivations for each term in \cref{sup_sec: derive bending force}.

There are four fundamental geometric vectors on halfedges that comprise the bending force:
the mean curvature vector (see \cref{fig: energy and force}B for indices),
\begin{equation}
    \int 2\vH_{ij} = \frac{1}{2} \left (\int 2 \vH_{i(jk)} + \int 2 \vH_{i(jl)} \right);
\end{equation}
the Gaussian curvature vector, 
\begin{align}
    \int \vK_{ij} = \frac{1}{2}  \varphi_{ij}  \nablari l_{ij};
\end{align}
and the two Schlafli vectors,
\begin{align}
    \int \vS_{ij,1} = \frac{1}{2} l_{ij} \nablari \varphi_{ij}, \quad \quad \int \vS_{ij,2} =  \frac{1}{2} \left( l_{jk} \nablari \varphi_{jk} + l_{jl}\nablari \varphi_{jl}  + l_{ji} \nablari \varphi_{ji} \right),
\end{align}
which act to smooth the profile of local dihedral angles.
Note that the shape derivatives are taken with respect to different vertices (i.e., $\nablari$ or $\nabla_{\vr_j}$), such that the mean curvature $\int \vH_{ij}$, Gaussian curvature $\int \vK_{ij}$, and Schlafli vectors $\int \vS_{ij}$ are asymmetric under index permutation.
To account for the orientation we refer to them as \emph{halfedge} vector quantities on $\underline{e}_{ij}$ (\cref{sec: halfedge}).
A numerical comparison of the discrete geometric vector with their smooth counterparts is shown in \cref{SI_fig: ptwise magnitude} and \cref{SI_fig: ptwise direction}.

The bending force $\int \vf^{\,b}_i$ (\cref{fig: energy and force}E -- bending) can be expressed as weights which are functions of input parameters multiplied by basic geometric measurements in scalar and vector form,  
\begin{align}\label{eqn: discrete bending force}
     \begin{split}
    \int \vf^{\,b}_i  = ~ \nearbyHalfedges & -\left[ \kappa_i (H_i - \bar{H}_i) + \kappa_j (H_j - \bar{H}_j) \right] \int \vK_{ij}\\
    & + \left[\frac{1}{3} \kappa_i (H_i - \bar{H}_i) (H_i + \bar{H}_i) + \frac{2}{3} \kappa_j (H_j - \bar{H}_j) (H_j + \bar{H}_j) \right] \int 2 \vH_{ij} \\
    & - \left[ \kappa_i (H_i - \bar{H}_i) \int \vS_{ij,1} + \kappa_j (H_j - \bar{H}_j) \int \vS_{ij,2}\right],
    \end{split}
\end{align}
which parallels the shape derivative of the smooth bending energy,
\begin{align}\label{eqn: smooth bending normal variation}
\begin{split}
    \nablar^\perp E_b &= \nablar^\perp \left [ \int_{\mathcal{M}} \kappa (H - \bar{H})^2 dA \right] \\
    &= \kappa \left [ 2(H - \bar{H})\left (  H^2 - K + \bar{H} H\right) +  \Delta_s (H - \bar{H})  \right] dA,
\end{split}
\end{align}
where $\nablar^\perp = \nablar \cdot \vn$ is the shape derivative in the surface normal direction.

Comparing the smooth-discrete expressions, we make a few observations:
\begin{itemize}
    \item The Schlafli vector terms, $\vS$, is the discrete analog of the smooth biharmonic term, $\Delta_s (H - \bar{H})$, the high-order local smoothing force. 
    The numerical comparison of these two terms, as well as results directly obtained using cotangent formula applied on pointwise scalar mean curvature, are covered in \cref{SI_fig: ptwise direction} and \cref{SI_fig: ptwise magnitude}.
    \item \cref{eqn: smooth bending normal variation} is normal component of the shape derivative of the bending energy; an additional tangential component is required if surface heterogeneity exists (e.g. $\kappa$ is not spatially uniform) ~\cite{steigmann1999fluid, guckenbergerTheoryAlgorithmsCompute2017}. 
    In contrast, the discrete shape derivative (\cref{eqn: discrete bending force}) is the total derivative in $\Real^3$, which includes both the tangential and normal components\footnote{In the smooth sense since there is no well-defined vertex normal direction in discrete geometry}. Depending on the extent and symmetry of the heterogeneity, discrete force can point in any direction in $\Real^3$. 
    \item The coefficients in \cref{eqn: discrete bending force} shows an intriguing pattern combining contribution from both $v_i$ and $v_j$.
    From a finite-difference approximation standpoint, this results in an approximation scheme for which a rigorous error analysis has not yet been conducted.
\end{itemize}

\subsubsection{Net force and the benefit of DDG}

Summing the force terms from each physics, we obtain a net force.
The expression for the total force is a function of geometric primitives and was derived using concepts from DDG without the need to introduce coordinates and use tensor algebra.
The numerical performance of these expressions are benchmarked for several scalar and vector measurements on smooth and discrete surfaces shown in \cref{SI_fig: ptwise magnitude}, \cref{SI_fig: ptwise direction}, and later discussed in \cref{sec: practical considerations for applying Mem3DG to biological problems}.
Also important, our resulting expressions allow the direct comparison of terms between smooth and discrete contexts.
By choosing definitions that preserve the chain of shape derivatives for the geometric vectors of interest (c.f., \cref{SI_fig: chain of variation}), we can preserve geometric intuition between smooth and discrete differential geometry theory~\cite{craneDISCRETEDIFFERENTIALGEOMETRY}.
With respect to the numerical performance, since the terms of the discrete shape equation are defined using values from neighborhoods around individual vertices the algorithms are efficient and straightforward to parallelize.
Owing to the local nature of the force evaluation, heterogeneities in material and other properties across the membrane are also straightforward to incorporate.

\subsection{Defining metrics for simulation and error quantification} \label{sec: defining metrics for convergence and comparison of forces}

For monitoring simulation progress, exactness of force calculations with respect to the discrete energy, and convergence studies of computed quantities upon mesh refinement we introduce the following norms.

\subsubsection{L2 norm}

From a \gls{pde} perspective, the vertex forces are also called the residual of the shape equation, whose solution represents the equilibrium solution.
The simulation is terminated when the residual is smaller than a user-specified threshold.
The rationale for using the $L_2$ norm is justified by perturbing the system configuration and conducting an expansion on the system energy, 
\begin{equation}\label{eqn: taylor expansion}
    \begin{split}
        E(\vbr + \epsilon \nabla E(\vbr)) & = E(\vbr)+ \epsilon \langle \nabla E(\vbr), \nabla E(\vbr) \rangle +\mathcal{O}\left(\epsilon^{2}\right)\\
        & = E(\vbr) + \epsilon \norm*{\int \vbf}_{L_2}^2+\mathcal{O}\left(\epsilon^{2}\right),
    \end{split}
\end{equation}
where we refer the inner product of the force matrix as the $L_2$ norm of the forces. 
Computationally, this is equivalent to the standard Frobenius matrix $L_2$ norm, 
\begin{equation}\label{eqn: L2 norm}
    \norm*{\int \vbf}_{L_2} = \sqrt{\mathrm{trace}\left(\int \vbf^\top \int \vbf\right)}.
\end{equation}
Using the $L_2$ norm and \cref{eqn: taylor expansion}, we can perform a numerical validation of the exactness of the discrete force calculation with respect to the discrete energy.
We expect the force to approximate the energy up to \nth{2} order with respect to the size of a perturbation.
This validation will be further elaborated in \cref{sec: protein variation}. 

\subsubsection{L1 norm}

A scale-invariant $L_1$ norm is well-suited to quantify the magnitude of the error on varying domain size and mesh resolution.  
Given a vertexwise local scalar measurement, $\int \ba$, or a vector measurement, $\int \vba$, and their reference values, $\int \bar{\ba}$, and $\int \bar{\vba}$,
\begin{equation}\label{eqn: L1 norm}
    \begin{split}
        \norm*{\int \ba}_{L_1} &= \frac{\sum_{v_i} |\int a_i - \int \bar{a}_i|}{A}  \\ 
        \norm*{\int \vba}_{L_1} &= \frac{\sum_{v_i} \|\int \va_i - \int \bar{\va}_i\|_{L_2}}{A},
    \end{split}
\end{equation}
where the normalizing factor, the total surface area \(A\), is used to obtain a pointwise estimate of the error.
The $L_1$ norm is applied in the local comparison of discrete and smooth measurements, where we further elaborate in \cref{sec: practical considerations for applying Mem3DG to biological problems}.

\subsection{Software implementation -- Mem3DG} \label{sec: software implementation}

\begin{figure}[htbp]
    \centering
    \includegraphics[width= \linewidth ]{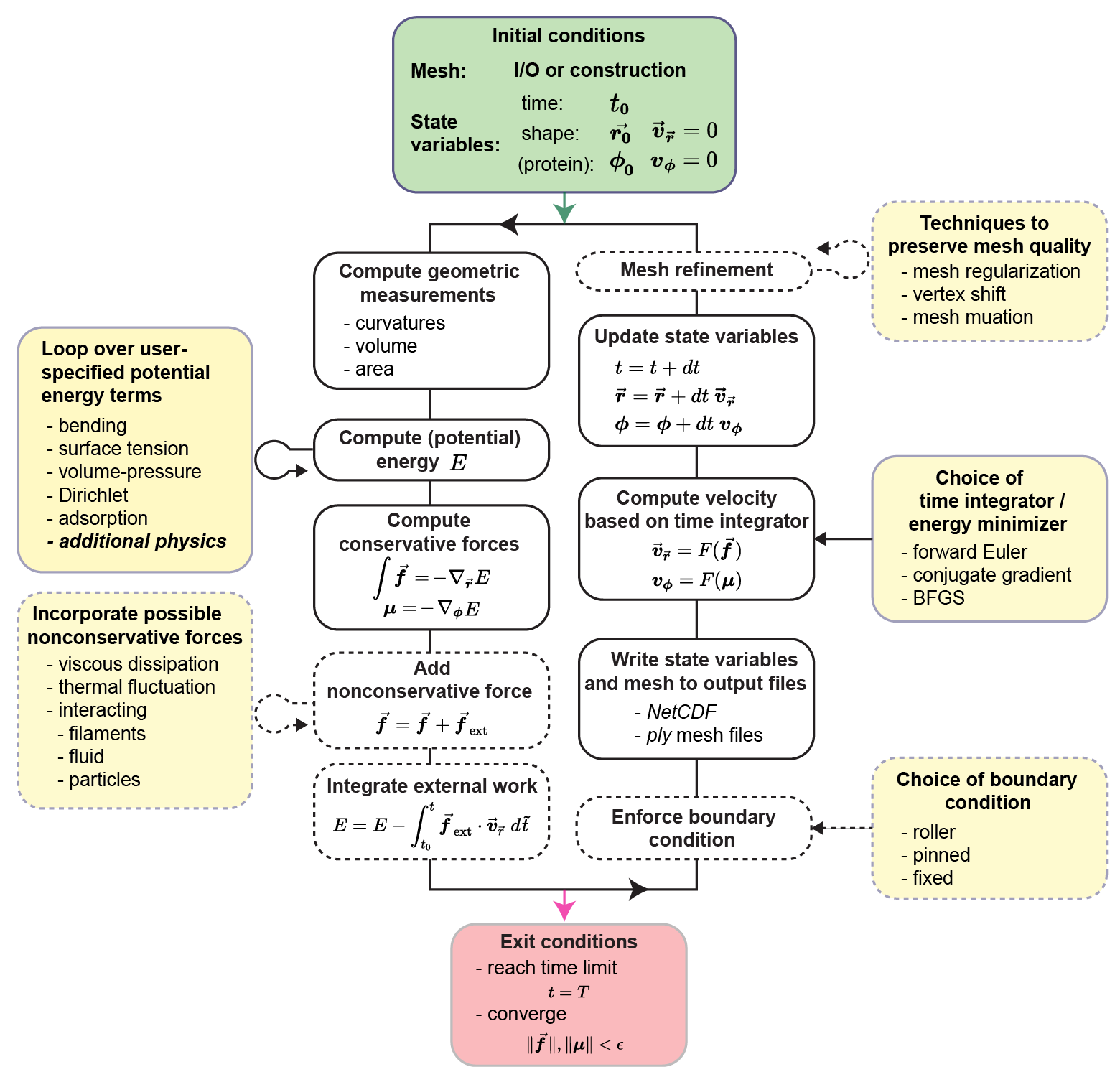}
    \caption{
        Overview of data flow within \mem3dg. 
        The user provides \mem3dg with an initial condition in the form of a triangulated mesh and vertexwise state and kinematic variables (green box). 
        The main loop (black loop) of \mem3dg evaluates the discrete energy and forces, and propagates the trajectory, among other supporting steps. 
        Modules in dashed lines are optional depending on whether the system mesh has boundaries and if external forces are specified. 
        User-specified options and possible extensions of \mem3dg to accommodate various physics are highlighted in yellow boxes.
        \mem3dg automatically exits the simulation when the system converges or the maximum time step is reached.
    }
    \label{fig: flowchart}
\end{figure}

Along with the theoretical developments, we have developed an accompanying software implementation written in C++ called \mem3dg.
Our goal in developing this software is to enable the easy use and application of the corresponding theory developed above to biological problems of interest.

\mem3dg is a library that contains several components to support this goal.
\cref{fig: flowchart} provides a synopsis of \mem3dg.
The input to \mem3dg includes a triangulated mesh with its coordinate $\vec{r}$ embedded in $\Real^3$.
Users can choose to use \mem3dg to construct idealized meshes (e.g., icosphere, cylinder, or flat hexagonal patch) as an input or to read in meshes from several common mesh formats. 
Meshes are stored and manipulated in \mem3dg using the half-edge data structure provided by Geometry Central ~\cite{geometrycentral}.
The supported input file formats are those which are readable by \texttt{hapPLY} and Geometry Central ~\cite{happly,geometrycentral}.
Once a mesh and parameters are loaded, \mem3dg can evaluate the discrete energy and forces of the system.
\mem3dg adopts a modular design that facilitates the use of different energy and force components and has utilities which help the user to specify the physics and governing parameters. 
\mem3dg also supports local system simulations where the input mesh has boundaries.
Additional details about the supported boundary conditions is given in \cref{sec: prescribing boundary conditions with force masking}.

To perform energy minimization and time integration of the system, various schemes have been implemented.
These schemes are described in \cref{sec: time integration and energy minimization}.
As discussed further in \cref{sec: practical considerations for applying Mem3DG to biological problems}, when a user wishes to use \mem3dg to represent complex biological membrane geometries, additional care regarding the quality of the mesh is necessary.
\mem3dg includes algorithms for basic mesh regularization and remeshing which can be toggled by the user to support their applications.
The simulation terminates when it reaches the time limit or the system reaches equilibrium, whose criteria is determined using the energy $L_2$ norm introduced in \cref{sec: defining metrics for convergence and comparison of forces}.
A user can choose between several formats to output a trajectory over time or the configuration of the local minima from \mem3dg.
In addition to the mesh outputs supported by Geometry Central, we have also developed a scheme for outputting mesh trajectories in NetCDF format ~\cite{rew1990netcdf}.
\mem3dg can read and visualize the output trajectories and mesh configurations using Geometry Central and Polyscope ~\cite{geometrycentral,polyscope}.

For rapid prototyping and enumeration of simulation conditions, we have also developed a Python API called \texttt{PyMem3DG}.
The functionality in C++ is exposed in Python using bindings from \texttt{pybind11} ~\cite{pybind11}.
Illustrative examples of using both \mem3dg and \texttt{PyMem3DG} are provided in the online tutorials.
For the experiments discussed in this work, all of the simulations were performed using \texttt{PyMem3DG} and the accompanying code and initial configurations are on GitHub: \url{https://github.com/RangamaniLabUCSD/Mem3DG}.

\subsubsection{Defining properties of a membrane reservoir for systems with open boundaries}
\label{sec: a unifying parametrization of the volume and area in local and global simulations}

To facilitate correspondence with wet experiments and to support the reduction of computational cost, it is possible to construct systems using meshes with open-boundaries in \mem3dg.
For example, when modeling the formation of a small endocytic bud from a large cell, the deformation is small compared to the broader system.
If we assume that the bulk of the cell is invariant with respect to bud formation, the computational burden can be reduced by modeling only the local deformation; we can assume that the modeled patch is attached to an implicit membrane reservoir.
To define this coupled system, the \textit{constant} area ($A_{r}$) and volume ($V_{r}$) of the reservoir must also be provided.
The total area and volume of the broader system is given by $A = A_{\mathrm{patch}} + A_{r}$, and $V = V_{\mathrm{patch}} + V_{r}$, where $A_{\mathrm{patch}}$ and $V_{\mathrm{patch}}$ are area and ``enclosed volume'' of the mesh patch respectively.
In our models, we enforce that all elements of a boundary loop are on the same plane; this way $V_{\mathrm{patch}}$ can be unambiguously defined as the enclosed volume when each boundary loop is closed by a planar sheet.
The capability to model systems attached to a reservoir reduces the modeled degrees of freedom while enabling intuitive physics to simplify the process of mimicking experimental conditions using \mem3dg.

\subsubsection{Prescribing boundary conditions with force masking}
\label{sec: prescribing boundary conditions with force masking}

\mem3dg supports modeling membranes with and without boundaries: a sphere (with no boundaries), a disk (with 1 boundary), and a open cylinder (with 2 boundaries).
For systems without boundaries, since the force calculation of the scheme is exact, the forces will not introduce artificial rigid body motions, as was also noted by \textcite{bianBendingModelsLipid2020}.
To study system with boundaries, \mem3dg currently supports three types of boundary conditions:
\begin{itemize}
    \item \textbf{Roller}, where the movement of boundary vertices is restricted along a given direction or plane.
    \item \textbf{Pinned}, where the position of boundary vertices are pinned while the curvature is allowed to vary.
    \item \textbf{Fixed}, where both the position and the boundary curvature are fixed for vertices on the boundary. 
\end{itemize}
The different boundary conditions are achieved by masking the elements of the force matrix corresponding to the boundary vertices and some neighborhood.
For example, to apply roller boundary conditions, we mask the Z--component of the force on the boundary vertices, therefore constraining their movement to the X--Y plane; pinned boundary conditions mask all force components for the boundary vertices to fix their position; fixed boundary conditions mask all force components for the outermost three layers to fix both their position and curvature.

\subsubsection{Time integration and energy minimization} \label{sec: time integration and energy minimization}

In this work, we use the forward Euler algorithm to integrate the system dynamics and the nonlinear conjugate gradient method to solve for equilibrium conditions. 
Both solvers are complemented by a backtracking line search algorithm, which satisfies Wolfe conditions to support adaptive time-stepping and robust minimization ~\cite{nocedalNumericalOptimization2006}.

The forward Euler scheme was chosen as the simplest dynamical propagator; physically it represents over-damped conditions where the environment of the membrane is too viscous for the system to carry any inertia.
Mathematically the physics is described by,
\begin{align}
    \dot{\vbr} = \frac{1}{\xi} \int \vbf = \frac{1}{\xi} \int (\vbf^{b} + \vbf^{s} + \vbf^p),
\end{align}
where $\xi$ is the drag coefficient.
From an optimization perspective, forward Euler is equivalent to the gradient descent method for minimizing an objective function, which is the discrete energy in our case.

A second propagator is the nonlinear conjugate gradient method for locally minimizing the discrete energy to yield the equilibrium shape of the membrane. 
Since the system is nonlinear we periodically perform forward Euler (gradient descent) steps after several conjugate gradient steps.
This approach of iterating between conjugate gradient and gradient descent steps is commonplace in the literature for solving nonlinear systems~\cite{nocedalNumericalOptimization2006}. 

We note that other time integrators and energy minimizers are also compatible with \mem3dg.
Included in the software are reference implementations of Velocity Verlet integration (for symplectic time integration), and Limited-memory Broyden–Fletcher–Goldfarb–Shanno algorithm (L-BFGS, a quasi-Newton method to the equilibrium shape for large scale problems where fast computation is needed).
We do not discuss these additional solvers in this work.

\subsubsection{Practical considerations for applying Mem3DG to biological problems} \label{sec: practical considerations for applying Mem3DG to biological problems}

As we have noted above, in the \gls{ddg} perspective, the mesh \textit{is} the geometry and thus the formulation of the discrete forces and energies is exact.
There are therefore very few restrictions on the resolution and quality of the input mesh.
However, in biophysics, we often consider biological membranes as smooth systems.
We expect that many users of \mem3dg may wish to approximate a smooth system using our discrete model.
In doing so, they make an implicit assumption that such an approximation is reasonable. 
Although the relationships between geometric objects and their shape are preserved between the smooth and discrete contexts, our ability to approximate a smooth problem with a discrete mesh is not guaranteed.
Similar to finite differences and \gls{fem}, additional constraints on mesh quality and resolution must be imposed.
To verify and understand the limitations of the assumption that the discrete mesh is the geometry and includes all of the geometric information, we numerically test the convergence of the discrete quantities under variation of resolution on an oblate spheroid mesh. 
The additional details regarding these numerical experiments are presented in \cref{sec: spheroid benchmark}.

\begin{figure}[htbp]
    \centering
    \includegraphics[width= 6in ]{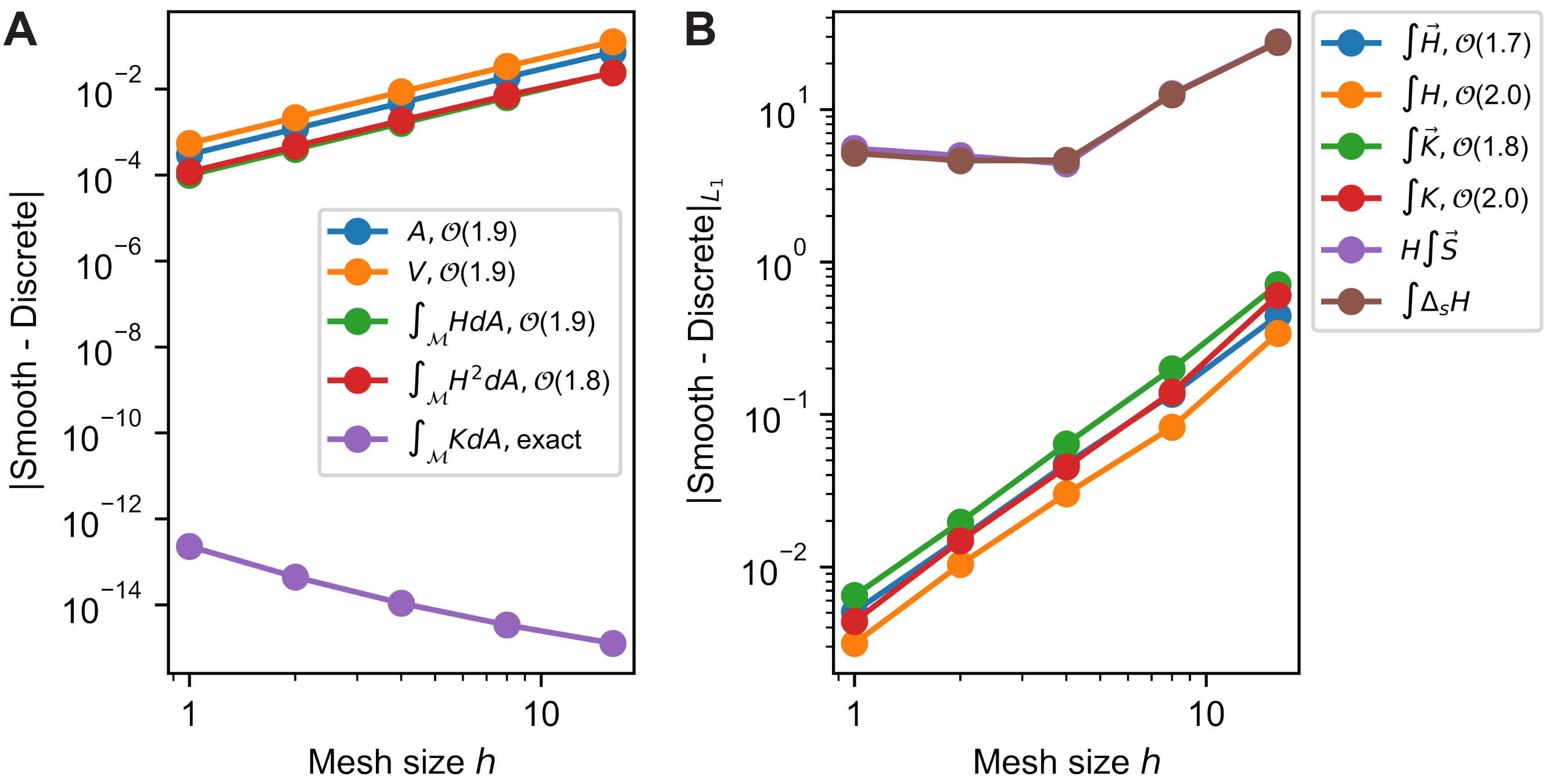}
    \caption{Comparison of discrete quantities with their smooth counterparts on spheroid shape. A) Convergence plot of global quantities, including total area, volume, mean curvature (squared), and Gaussian curvature, and B) Convergence plot of $L_1$ norm of scalar and vector local quantities, including the mean curvature, Gaussian curvature, and the biharmonic term.}
    \label{fig: spatial convergence}
\end{figure}

Setting the characteristic length scale of the finest mesh to be $h=1$, as the mesh coarsens (i.e., mesh size increases) $h$ increases.
\cref{fig: spatial convergence} shows the scaling relationship of the deviation in magnitude between the smooth and discrete quantities.
\cref{fig: spatial convergence}A shows the convergence property of global measurements that determines the energy (\cref{eqn: smooth free energy} and \cref{eqn: discrete free energy}), including the total area, $A$, enclosed volume, $V$, and total Gaussian curvature and mean curvature (squared), $\int_{\mathcal{M}} K dA$, $\int_{\mathcal{M}} H dA$ and $\int_{\mathcal{M}} H^2 dA$, respectively.
Except for the total Gaussian curvature being an exact topological invariant, all integrated quantities exhibit approximately \nth{2} order convergence rate.

We acknowledge that convergence of global measurements does not imply that local measurements will also converge.
To validate the convergence of local measurements, which determines the convergence of local forces on the membrane (e.g., \cref{eqn: discrete osmotic force}, \cref{eqn: discrete capillary force} and \cref{eqn: discrete bending force}), we utilize the $L_1$ norm (\cref{eqn: L1 norm}) to evaluate the deviation of local quantities from their smooth counterparts. 
\cref{fig: spatial convergence}B shows the local convergence plot. 
Similar to their global counterparts, local scalar mean and Gaussian curvature, $\int H$, and $\int K$, converge at $\mathcal{O}(h^2)$. 
\cref{fig: spatial convergence}B also shows the convergence of vector quantities, which not only contribute to the magnitude of the force but also set the direction of the force.
The test shows that most vector quantities converge slightly slower than their scalar counterparts.
Two terms exhibit poor convergence, the Schlafli vector term in \cref{eqn: discrete bending force}, $H\int S$, and a scalar counterpart, $\int \Delta_s H$.
The latter term corresponds to the direct application of the cotangent Laplacian (\cref{eqn: discrete capillary force}) to the pointwise scalar mean curvature field; this approach is not used in our framework but is common in the literature\cite{guckenbergerTheoryAlgorithmsCompute2017}.
Both non-convergent expressions are discrete representations of the biharmonic term, $\Delta_s H$, which have been noted to be sensitive to noise of vertex coordinates in the prior literature~\cite{guckenbergerBendingAlgorithmsSoft2016}.
Recall that the biharmonic term is the fourth-order derivative of the embedded coordinates.
Although the traditional approximation theories suggest that higher-order derivatives often exhibit slower rates of convergence~\cite{hughes2012finite}, to the best of our knowledge, there is not yet a rigorous study that connects \gls{ddg} with an approximation theory.
Nevertheless, we anticipate that similar principles hold.
Two spatial plots comparing local measurements between smooth and discrete contexts are provided in the appendix (\cref{SI_fig: ptwise magnitude} and \cref{SI_fig: ptwise direction}); each test is conducted using the finest mesh size ($h = 1$). 
Based on this numerical validation, we conclude that the computation of energy converges with a \nth{2} order rate (\cref{fig: spatial convergence}A). 
While most components of the forces converge, the biharmonic term remains a limiting factor.

One other practical consideration for our models is that the Helfrich Hamiltonian, matching the in-plane fluidity of biological membranes, has no resistance to shearing. 
Without additional constraints, the mesh is susceptible to shearing motions which can deteriorate mesh quality in some conditions~\cite{vasanMechanicalModelReveals2020}.
This can degrade the implicit assumption that the discrete mesh represents a smooth geometry.
To ensure that such an approximation can remain valid throughout a trajectory, we have incorporated algorithmic solutions to adaptively maintain an isotropically well-resolved discrete geometry. 
This is achieved by two operations: 
1) mesh regularization using local force constraints which are common in finite element methods ~\cite{sauerStabilizedFiniteElement2017, vasanMechanicalModelReveals2020, auddyaBiomembranesUndergoComplex2021, fengFiniteElementModeling2006} \cref{sec: mesh regularization}.
and 2) mesh mutations such as decimating, flipping, and collapsing edges.
Mesh mutations are also a common practice to cope with deterioration in other mesh simulations which use a Monte Carlo integration~\cite{krollConformationFluidMembranes,atilganShapeTransitionsLipid2007a,bahramiFormationStabilityLipid2017,noguchiShapeTransitionsFluid2005a, jieNumericalObservationNonaxisymmetric1998, bianBendingModelsLipid2020,tachikawaGolgiApparatusSelforganizes2017,brakkeSurfaceEvolver1992}.
The algorithms for mesh regularization and mutation are further described in \cref{sup_sec: mesh regularization and mesh mutation}.

\section{Results and Discussion}

To further validate the method and to provide a sense of how \mem3dg can be used and extended to solve more complex physics, we apply \mem3dg to a sequence of examples with increasing complexity. 
First, we model well-studied systems with homogeneous membrane conditions.
We show that \mem3dg is capable of reproducing the classical solutions without imposing the axisymmetric constraint commonly adopted by other solvers. 
The later examples set a blueprint for extending and modifying \mem3dg for particular systems of interest.
We introduce new energy and corresponding force terms to expand the formulation for complex systems of interest. 
We emphasize that the goal of these examples is to illustrate the generality of the theory and software and to outline specific steps for future extensions; we do not perform rigorous experimental comparisons nor extract scientific insights.
Additional care must be taken to mimic specific biological experiments for model validity, which is left for future work.

Each of the following sections considers a different class of membrane biophysics problem of increasing complexity in the coupling of the in-plane protein density parameter, $\phi \in [0, 1]$.
To mimic the various influences protein-lipid interactions can have on the bilayer, the protein density can be set to influence membrane properties such as the spontaneous curvature, $\bar{H}(\phi)$, and bending rigidity, $\kappa(\phi)$.
More complex phenomena such as the production of in-plane interfacial forces from membrane-protein phase separation\cite{liu2006endocytic, julicher1996shape, saleemBalanceMembraneElasticity2015} can also be modeled.
In our final proof of concept, we extend \mem3dg to support full mechanochemical dynamics, where proteins can mobilize in- and out-of-plane through adsorption and lateral diffusion, based on its coupling with membrane material properties and shape transformation.
These scenarios highlight the relative ease of extending \mem3dg with additional physics and the potential utility to simulate realistic experimental scenarios. 
Note that for all of the examples, unless otherwise specified, the bending rigidity of membrane, $\kappa$, is assumed to be the rigidity of a bare membrane, $\kappa_b = \SI{8.22e-5}{\um \nano \newton}$. 
Despite the superior performance of the nonlinear conjugate gradient method in finding an energy minimizing configuration, to maintain both static and dynamic interpretability, we perform all simulations using a forward Euler integrator unless otherwise noted.
All simulations presented in this work were conducted on a standard workstation with Intel Xeon processors.
Although the numerical algorithms are amenable to parallelization, \mem3dg is currently a single-threaded program. 
Using a single core, the simulations here complete in minutes and up to two hours.

\subsection{Modeling spherical and cylindrical membranes with homogeneous physical properties} \label{sec: modeling spherical and cylindrical membranes with homogeneous physical properties}

\begin{figure}[htbp]
    \centering
    \includegraphics[width=6 in]{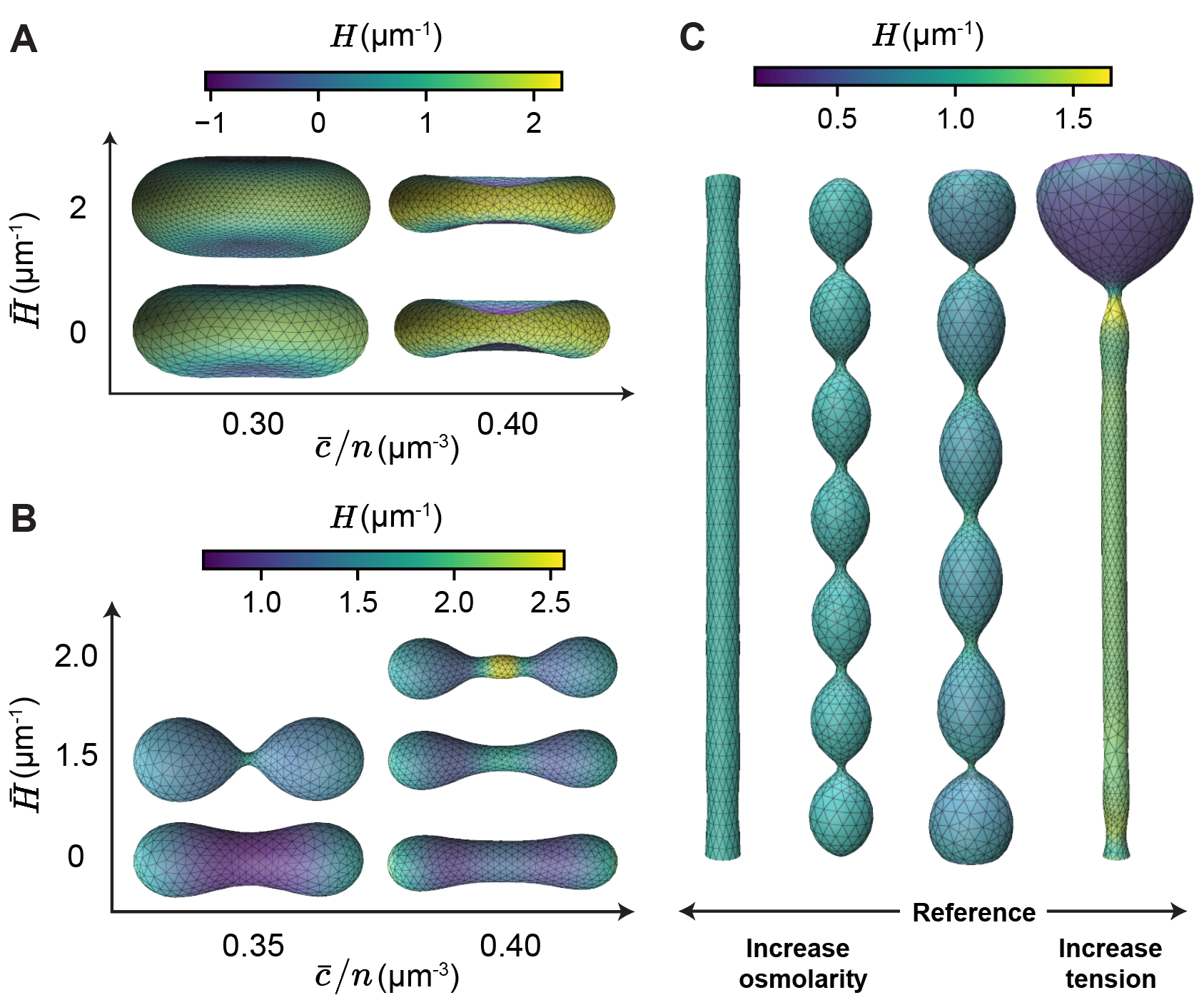}
    \caption{Recover typical equilibrium shapes of membranes with homogeneous material properties.
    A-B) Equilibrium solutions under different osmolarity ($\bar{c}$) and spontaneous curvature ($\bar{H}$) conditions, with initial condition of A) Oblate spheroid and B) Prolate spheroid.
    We vary the osmolarity by adjusting the concentration of the ambient solution, $\bar{c}$, while holding the enclosed amount of solute, $n$, constant.
    C) Equilibrium solutions of a tubular membrane structure under variations in osmolarity and surface tension.}
    \label{fig: homogeneous}
\end{figure} 

\begin{movie}
    \caption{Animated time series simulation of \cref{fig: homogeneous}C--reference. The color map indicates the local pointwise mean curvature, $H$, and $t$ represents the numerical time. Available on GitHub: \url{https://github.com/RangamaniLabUCSD/2020-Mem3DG-Applications/blob/master/examples/videos/compressed/pearl_control_1.mp4}}
    \label{mov: pearl control}
\end{movie}
\begin{movie}
    \caption{Animated time series simulation of the unduloid shape shown in \cref{fig: homogeneous}C--medium osmolarity. The color map indicates the local pointwise mean curvature, $H$, and $t$ represents the numerical time. Available on GitHub: \url{https://github.com/RangamaniLabUCSD/2020-Mem3DG-Applications/blob/master/examples/videos/compressed/pearl_pearl_1.mp4}}
    \label{mov: pearl pearl}
\end{movie}
\begin{movie}
    \caption{Animated time series simulation of the beads-on-a-string structure shown in \cref{fig: homogeneous}C--high tension. Disruption of the intermediate metastable state, a symmetric structure with two beads shown in \cref{SI_fig: metastable two beads}, occurs. The color map indicates the local pointwise mean curvature, $H$, and $t$ represents the numerical time. Available on GitHub: \url{https://github.com/RangamaniLabUCSD/2020-Mem3DG-Applications/blob/master/examples/videos/compressed/pearl_bead_1.mp4}}
    \label{mov: pearl bead}
\end{movie}

We begin our examples by using \mem3dg to find the equilibrium shapes of membranes with homogeneous protein density, $\phi$.
We ask, given an initial membrane configuration with uniform bending modulus and spontaneous curvature, what are the minimizers of the system energy?
The answers are the classical equilibrium solutions to the shape equation obtained analytically~\cite{naitoNewSolutionsHelfrich1995}, and numerically using many methods with different assumptions~\cite{helfrich1973elastic,deuling1976red}.
We will show solutions obtained using \mem3dg with topologies of sphere and tube (\cref{fig: homogeneous}).
These geometries were selected not only because of their potential for comparison with the legacy literature but also because they are reminiscent of biological membranous structures such as red blood cell~\cite{EVANS1972335,AlimohamadiEtAl2019}, cell-cell tunneling and tethering~\cite{gerdes2008intercellular,PearceEtAl2020,alimohamadi2020modeling}, neuron beading~\cite{datarRolesMicrotubulesMembrane2019, pullarkatOsmoticallyDrivenShape2006}, among other biological processes.

Starting with closed topological spheres, \cref{fig: homogeneous}A and B shows the typical equilibrium shapes under osmotic stress while the surface area is conserved.
The preferred area of the vesicle, $\bar{A} = \SI{4 \pi}{\micro\meter\squared}$, represents a sphere of radius \SI{1}{\micro\meter}.
This constraint is achieved by prescribing a large stretching modulus, $K_A$, such that the areal strain, $(A - \bar{A}) / A$, is less than one percent.
The strength constant of osmotic pressure, $K_V$ is set to be \SI{0.1}{\um \nano \newton}.
Initializing the simulations from an oblate spheroid, as the osmolarity increases (e.g., the normalized ambient solution, $\bar{c}/n$), we recover the well-known biconcave red blood cell shape~\cite{deuling1976red,Evans1974} (\cref{fig: homogeneous}A).
The vesicle adopts a more convex configuration as we increase the spontaneous curvature, indicating an overall increase in its mean curvature with the concomitant decrease of areas with negative mean curvature (the dimple regions).
In contrast, starting from a prolate spheroid, as the spontaneous curvature increases, the vesicle adopts a dumbbell configuration as the energetically preferred state (\cref{fig: homogeneous}B).
The size of the beads on the dumbbell is governed by the osmolarity, $\bar{c}/n$.
These trends with respect to the variations of the spontaneous curvature and osmolarity are consistent with the analytical and numerical results in the broader literature~\cite{bianBendingModelsLipid2020,naitoNewSolutionsHelfrich1995}.
Qualitatively the predicted geometries of closed vesicles with homogeneous spontaneous curvature match the predictions of a detailed benchmark of mesh-based methods performed by \textcite{bianBendingModelsLipid2020}.

We also modeled the shapes of membranes starting from an open cylinder configuration under different osmotic and surface tension conditions (\cref{fig: homogeneous}C).
This problem is related to a well-studied phenomenon called the Plateau-Rayleigh instability~\cite{plateau1873,rayleigh1878}.
The Plateau-Rayleigh instability describes how surface tension  breaks up a falling stream of fluid into liquid droplets.
Compared with a liquid stream, lipid membrane provides additional resistance against the instability due to its rigidity. 
\textcite{zhong-canBendingEnergyVesicle} obtain stability regimes as a function of membrane bending rigidity and spontaneous curvature using the spectral stability analysis~\cite{zhong-canBendingEnergyVesicle}.
Though osmotic pressure is often reported as an important cause of morphological instability~\cite{pullarkatOsmoticallyDrivenShape2006, bar1994instability,sanbornTransientPearlingVesiculation2013}, the effect of osmotic pressure is difficult to isolate in wet experiments because change to osmolarity affects the surface tension, which is a key driver of the instability.
In our simulations, the two effects are decoupled, making the investigation of individual contributions to the morphology possible.
All shapes in \cref{fig: homogeneous}C evolve from the initial tubular mesh with radius of \SI{1}{\um} and axial length of \SI{19.9}{\um}, under a constant spontaneous curvature of \SI{1}{\per\um}.
These simulations are set up as local models (c.f., \cref{sec: a unifying parametrization of the volume and area in local and global simulations}) where the explicit mesh is assumed to be coupled to a membrane reservoir.
Additional geometric information defining the membrane reservoir and boundary conditions are required to initialize the the local model. 
The tubular structure is considered to be a cylinder that connects two otherwise detached domains (e.g., membrane reservoirs), which combined have a total reservoir volume, $V_r = \SI{4.19}{\um \cubed}$. 
The strength of osmotic pressure, $K_V$, is set to be \SI{0.01}{\um \nano \newton}. 
To isolate the effect of osmotic pressure and surface tension on the morphology, we prescribe a specific surface tension which we assume to be invariant with respect to changes to the surface area. 
On the two boundary loops of the mesh we apply roller boundary conditions, which restrict the movement of boundary vertices in the axial direction.
The length of the tube is thus constrained to be \SI{19.9}{\um} while the radius of the tube including the boundaries is free to shrink or expand.

As the osmolarity increases from the reference condition ($\bar{c}/n = \SI{0.022}{\per \um \cubed }$) (\cref{mov: pearl control}),
we obtain solutions such as the unduloid (or pearled) structure at $\bar{c}/n = \SI{0.030}{\per \um \cubed }$ (\cref{mov: pearl pearl}), and tube at $\bar{c}/n = \SI{0.051}{\per \um \cubed }$, which follow the trends from both analytical~\cite{naitoNewSolutionsHelfrich1995} and experimental observations~\cite{bar1994instability, pullarkatOsmoticallyDrivenShape2006,YuanEtAl2021}.
As we increase the surface tension from the reference condition ($\lambda = \SI{1e-7}{\nano \newton \per \um}$) to a tension-dominated regime ($\lambda = \SI{1e-4}{\nano \newton \per \um}$), we obtain the beads-on-a-string structure which minimizes the surface-to-volume ratio (\cref{mov: pearl bead}).
The formation of beads-on-a-string is an interesting configuration which has been identified in biological membranes and other systems~\cite{datarRolesMicrotubulesMembrane2019, pullarkatOsmoticallyDrivenShape2006}. 
Note that our simulations revealed a symmetric metastable state where two large beads forms at either end (\cref{SI_fig: metastable two beads}), connected by a thin tube, prior to adopting the asymmetric conformation shown in \cref{fig: homogeneous}C.
We believe that discretization artifacts such as mesh mutations acts as a perturbation to break the symmetry of the metastable intermediate and transition the membrane to a single bead configuration.

These examples with uniform spontaneous curvature profile proves the ability of \mem3dg to reproduce the expected classical solutions for spherical and tubular membrane geometries.
Note that no axisymmetric constraint is imposed in these simulations.
\mem3dg solves the system in full three dimensions and the symmetrical configurations is due to the problem physics.
The ability to adapt changing and complex curvatures of the membrane using discrete mesh is achieved using mesh mutation and other manipulations within solver steps.
For example, the pinched neck regions of the tubes are automatically decimated with finer triangles than other regions of the mesh.
For a global closed membrane simulation such as in \cref{fig: homogeneous}A, B, we do not remove any rigid body motions from the system;
Since the forces from \gls{ddg} are exact and we used the forward Euler integrator, no artificial rigid body motions are introduced throughout the simulation.
These examples show that that the derivation of the discrete energy and forces along with the software implementation are accurate and proceed to test \mem3dg with more complex examples.

\subsection{Modeling endocytic budding mechanisms}
\label{sec: robust endocytic budding mechanism}

\begin{figure}[htbp]
    \centering
    \includegraphics[width=\linewidth]{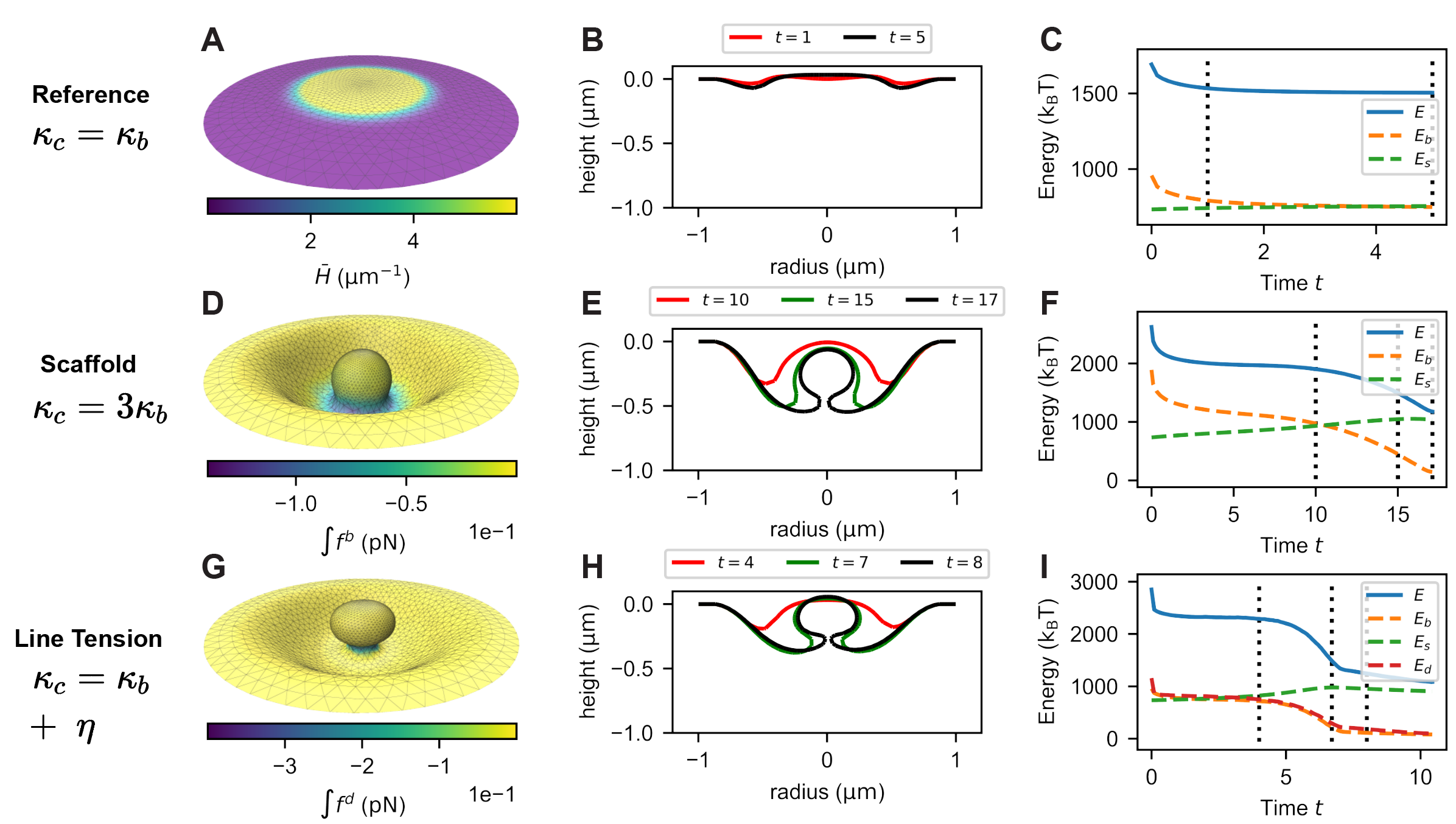}
    \caption{Budding dynamics by robust mechanisms of protein scaffolding and interfacial line tension constriction. A-C) Control group, D-F) Bending-driven scaffolding mechanism, and G-I) Interfacial line tension assisted budding. A) Spontaneous curvature distribution, $\bar{H}$, on initially flat patch. D) Normal projection of the bending force at $T = 15$. G) Normal projection of the line tension force at $T = 7$. B, E, H) Shape evolution through time-series snapshots of the Y-Z cross-sections of the membrane, corresponding to the vertical dash lines in C, F, I) Trajectory plots of system energy and its competing components.}
    \label{fig: budding dynamics}
\end{figure}

\begin{movie}
    \caption{Animated time series simulation using the reference parameters, shown by \cref{fig: budding dynamics}A-C. The color map indicates the local pointwise spontaneous curvature and $t$ represents the numerical time. Available on GitHub: \url{https://github.com/RangamaniLabUCSD/2020-Mem3DG-Applications/blob/master/examples/videos/compressed/patch_control_1.mp4}}
    \label{mov: patch control}
\end{movie}
\begin{movie}
    \caption{Animated time series simulation of the bending-rigidity-driven budding, shown by \cref{fig: budding dynamics}D-F. The color map indicates the projection of the bending force onto the surface normal and $t$ represents the numerical time. Available on GitHub: \url{https://github.com/RangamaniLabUCSD/2020-Mem3DG-Applications/blob/master/examples/videos/compressed/patch_Kb_1.mp4}}
    \label{mov: patch Kb}
\end{movie}
\begin{movie}
    \caption{Animated time series simulation of budding driven by interfacial line tension, shown by \cref{fig: budding dynamics}G-I. The color map indicates the signed projection of the line tension force onto the surface normal and $t$ represents the numerical time. Available on GitHub: \url{https://github.com/RangamaniLabUCSD/2020-Mem3DG-Applications/blob/master/examples/videos/compressed/patch_eta_1.mp4}}
    \label{mov: patch eta}
\end{movie}

Our goal is to highlight the potential of \mem3dg and its associated framework for investigating mechanical phenomena relevant to cellular biology.
Endocytosis is a cellular process in which cells uptake cargo from the extracellular environment;  the transported material is engulfed by the cell membrane which then buds off to form a vesicle~\cite{McMahonEtAl2015}. 
Endocytosis occurs through various mechanisms including the clathrin-mediated endocytosis~\cite{McMahonEtAl2015,WatanabeEtAl2013}.
It has been shown that clathrin aggregates on the plasma membrane helping to deform the membrane and form a spherical bud~\cite{McMahonEtAl2015,AvinoamEtAl2015,saleemBalanceMembraneElasticity2015}. 
However, the specific mechanisms of how membrane-clathrin interactions facilitate membrane curvature generation remains unresolved.
While there is significant literature investigating the many proposed mechanisms, here we develop models to demonstrate the bud formation via spatially localized spontaneous curvature, combined with a line tension term arising from phase separations on the membrane\cite{baumgartImagingCoexistingFluid2003}.

We model endocytic budding on a circular patch with radius \SI{1}{\um} (a disc with one boundary loop).
We assume that the patch is a \emph{local} system which is coupled to a large vesicle (\cref{sec: a unifying parametrization of the volume and area in local and global simulations}).
A heterogeneous protein density, $\phi \in [0,1]$, is applied to mimic the distribution of clathrin and other scaffolding proteins.
Shown in \cref{fig: budding dynamics}A, the protein density is high ($\phi = 1$) towards the center of the a geodesic disk with radius \SI{0.5}{\um}) and decreases towards the boundaries ($\phi = 0$).
During simulation, the geodesic distance to the center of the patch is periodically computed using the heat method~\cite{craneHeatMethodDistance2017}.
Vertexwise $\phi$ is assigned based on stair step profile smoothed by the hyperbolic tangent function applied to the geodesic distance. 
Each experiment is initialized as a flat patch and the displacement of boundary vertices are restricted using a fixed boundary condition. 
Since the patch is viewed as a small piece within a larger closed vesicle reservoir, we assume that the surface tension is constant.

A common model to account for the preferential bending owing to protein-membrane interactions is through the spontaneous curvature; we assume $ \bar{H}(\phi) = \bar{H}_c \, \phi$, where $\bar{H}_c = \SI{6}{\per \um}$ is the spontaneous curvature imposed by the membrane protein coat.
Proteins such as clathrin are known to form stiff scaffolds on the membrane.
Similar to the spontaneous curvature, we can assume a linear relationship between bending rigidity and protein density, $ \kappa (\phi) = \kappa_b + \kappa_c \, \phi$, where constant $\kappa_b$ is the rigidity of the bare membrane, and $\kappa_c$ is additional rigidity of the protein scaffold.

Shown in \cref{fig: budding dynamics}A-C and \cref{mov: patch control}, is the control simulation where we set the contribution to the rigidity from protein to be the same as that of the raw membrane, $\kappa_c = \kappa_b$. 
\cref{fig: budding dynamics}A shows the initial flat configuration of the control experiment; the color bar shows the heterogeneous spontaneous curvature resulted from the prescribed protein density profile.
In the control experiment, the bending force is resisted by the surface tension (\cref{fig: budding dynamics}C) until, at the final frame in \cref{fig: budding dynamics}B ($t = 5$), the membrane reaches the equilibrium configuration where the surface tension cancels with the bending force. 
In a second model, we assume that the scaffolding proteins is much more rigid than the bare membrane, $\kappa_c = 3\kappa_b$.
\cref{fig: budding dynamics}D-F and \cref{mov: patch Kb} show the bud formation due to this increased protein scaffolding effect. 
The greater rigidity results in an increase of initial bending energy, which outcompetes the resistance from the surface tension (\cref{fig: budding dynamics}F). 
\cref{fig: budding dynamics}E shows the shape evolution from a flat patch to a successful bud with pinched neck. 
\cref{fig: budding dynamics}D shows the signed projection of the bending force onto the vertex normal, $\int f^b_i = \int \vf^b_i \cdot \vn_i$, at $T = 15$.\footnote{Outward-pointing angle-weighted normal; the same applies to the interfacial line tension.}
We can see an ``effective line tension'' driven by the heterogeneous spontaneous curvature which constricts the neck.
This phenomenon is theoretically explored in detail by \textcite{alimohamadi2018role}.

For our third model, based on the prior observations that protein phase separation on surfaces can lead to a line tension~\cite{baumgartImagingCoexistingFluid2003}, we incorporate a Ginzburg-Landau interfacial energy into the system,
\begin{equation} \label{eqn: dirichlet energy}
    E_d = \frac{1}{2} \sum_{f_{ijk}} \eta \int \| \nablat \, \phi \|^2_{ijk}   \rightarrow \frac{1}{2} \int_\mathcal{M} \eta \| \nablat \, \phi \|^2~ dA.
\end{equation}
where $\eta$, referred to as the Dirichlet energy constant, governs the strength of the energy penalty, and $\nablat \,\phi$ is the discrete surface gradient of protein density profile.
The term is similar to previous modelling efforts by \textcite{elliottModelingComputationTwo2010} and \textcite{ma2008viscous} using \gls{fem}; 
because we use the protein phase separation as a prior, we exclude the double-well term which models the thermodynamics of phase-separation, and incorporate only the Dirichlet energy component that penalizes the heterogeneity of membrane composition.

Defined as the slope of the linearly interpolation of $\phi$ on faces of the mesh, $f_{ijk}$, the discrete surface gradient of the protein density is,
\begin{align}
    \nablat ~\phi_i = \frac{1}{2 A_{ijk}}\sum_{\underline{e}_{i} \in N(f_{ijk})} \phi_i \ve_i^{~\perp},
\end{align}
where following illustration in \cref{fig: energy and force}C, $\vec{\underline{e}}_{i}$ is the vector aligned with the halfedge $\underline{e}_{i}$, with its length of $l_{i}$, and $(\cdot)^\perp$ represents a \ang{90} counterclockwise rotation in the plane of $f_{ijk}$.
The resulting line tension force $\int \vf^d$ is then the shape derivative of the Dirichlet energy, $\nablar E_d$, which acts to minimize the region with sharp heterogeneity.
The detailed derivation of the shape derivative is elaborated in \cref{sup_sec: derive line tension and diffusion as variations of Dirichlet energy}, where we follow the formulaic approach by taking geometric derivatives of basic mesh primitives shown in \cref{eqn: derivatives of angles}.
\cref{fig: budding dynamics}G-I and \cref{mov: patch eta} show the trajectory where we used control bending rigidity, $\kappa_c = \kappa_b$, and the additional interfacial line tension, $\eta = \SI{5e-4}{\um \nano \newton}$.
We find that the interfacial line tension, jointly with the bending force, lowers the system energy and help the formation of a spherical bud (\cref{fig: budding dynamics}I, H).
\cref{fig: budding dynamics}G shows the snapshot ($t = 7$) with the color map representing the signed normal projection of the interfacial line tension that acts to constrict the neck.
These examples of endocytic bud formation help to illustrate the utility of \mem3dg and the accompanying theoretical framework.
Since physical parameters are assigned on a per-vertex basis, it is straightforward to incorporate heterogeneity such as the nonuniform bending rigidity and spontaneous curvature.
In smooth theory and its derived discrete mesh models, when membrane is heterogeneous, it is required to decompose the force separately in normal and tangential direction~\cite{steigmann1999fluid, guckenbergerTheoryAlgorithmsCompute2017};
In contrast, the general derivation of the discrete bending force following the formalism of \gls{ddg} permits modeling membrane with heterogeneous material properties without any modification to its formulation (\cref{sec: forces from bending}).
The introduction of Dirichlet energy and line tension force serves to highlight the relative ease to extend the modelled physics.

\subsection{Protein aggregation on the realistic mesh of a dendritic spine}\label{sec: spine}

\begin{figure}[htbp]
    \centering
    \includegraphics[width= 6 in]{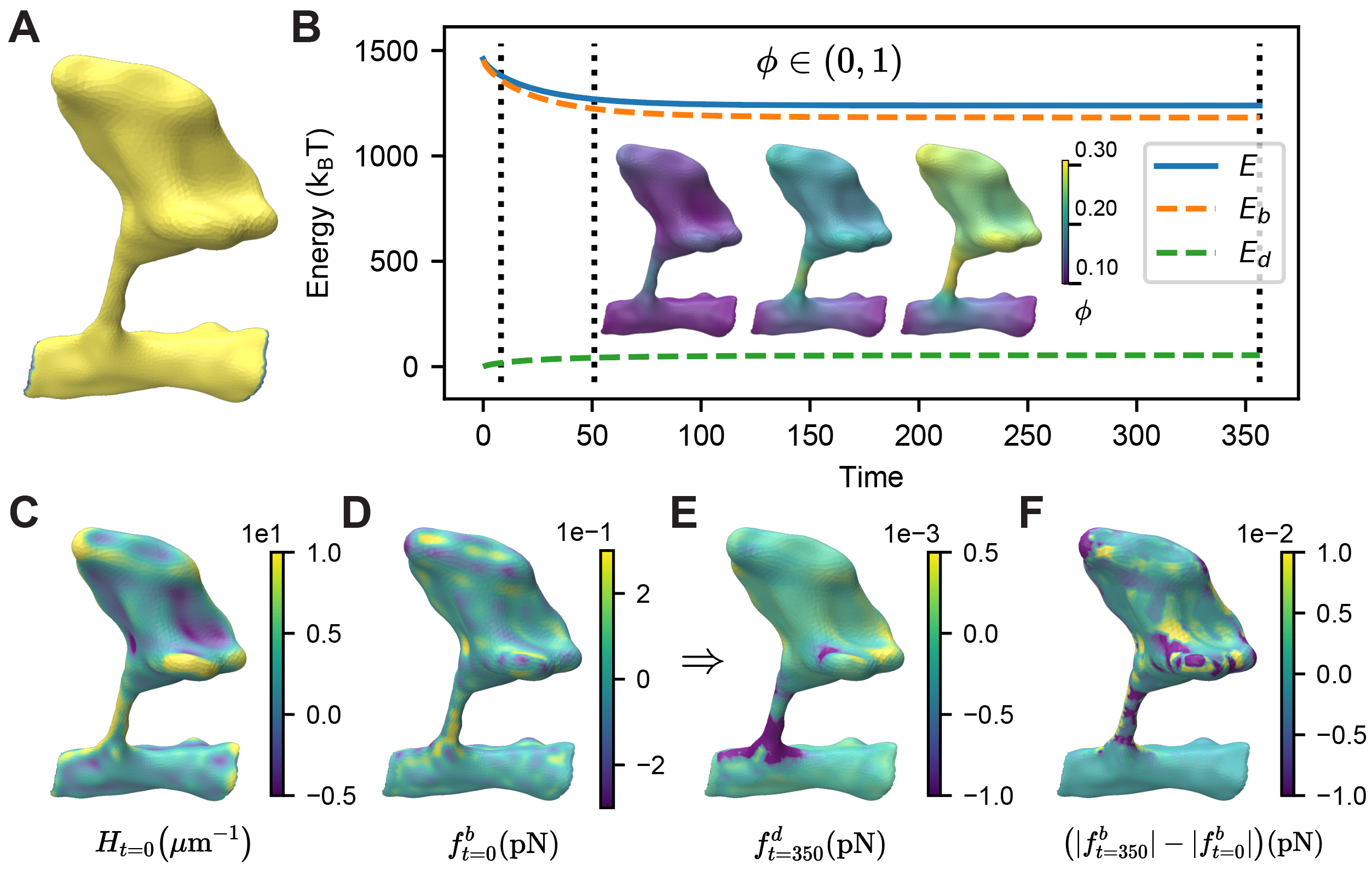}
    \caption{Protein aggregation on a realistic dendritic spine geometry. A) Mesh of the dendritic spine and its boundary elements. B) Trajectory of protein evolution and components of system energy. C) Mean curvature of the geometry. The normal component of D) the bending force at $t=0$, E) the line tension force produced by the equilibrium protein distribution, and F) the difference in the bending force profile produced by final protein distribution as opposed to the initial distribution.}
    \label{fig: spine}
\end{figure}
\begin{movie}
    \caption{Animated time series simulation of protein aggregation on a dendritic spine (\cref{fig: spine}B). The color map shows the order parameter for protein density, $\phi \in (0,1)$, and $t$ represents the numerical time. Available on GitHub: \url{https://github.com/RangamaniLabUCSD/2020-Mem3DG-Applications/blob/master/examples/videos/compressed/spine_1.mp4}}
    \label{mov: spine}
\end{movie}

\begin{table}[htbp]
    \caption{Parameters used in \cref{sec: spine}} \label{table: parameter in spine}
    \begin{center}
        \begin{tabular}
            {@{}cc@{}}
            Parameters  &Values   \\ \toprule
            $\phi_0 $ & 0.1 \\
            $\kappa_c $ & \SI{0}{\nano \newton \um} \\
            $\bar{H}_c $ & \SI{10}{\per \um} \\
            $B$     & \SI{3}{\per \nano \newton  \per \um  \per \second }  \\
            $\eta$  & \SI{0.01}{\um \nano \newton}  \\    
        \end{tabular}
    \end{center}
\end{table}

While the prior examples have focused on the mechanical response of the membrane given a bound protein distribution, we can also model the inverse problem.
Given the membrane shape, how do curvature-sensing proteins diffuse in the plane of the membrane and distribute over the domain?
And how does the resultant protein distribution influence the stresses of the system? 
To model the protein dynamics, we use three terms corresponding to protein binding, curvature sensitivity, and lateral diffusion.

To model the binding of proteins to the membrane, we assume that the energy of adsorption, $\varepsilon$, is constant and uniform across the surface such that the discrete adsorption energy is,
\begin{equation}\label{eqn: protein adsorption energy}
    E_{a} = \varepsilon \sum_i \int \phi_i,
\end{equation}
where $\phi_i$ is an order parameter representing the area density of protein at each vertex.
Taking the derivative with respect to $\phi$, referred to as the chemical derivative,
\begin{equation}
    \mu^a_{i} = -\nablap E_a  = -\int \varepsilon,
\end{equation}
we obtain the adsorption component of the chemical potential.
To account for protein curvature-sensitivity, we find the chemical potential of the bending energy,
\begin{equation}
    \label{eqn: bending chemical potential}
    \mu^b_{i} = -\nablap E_b =  \int [ 2 \kappa_i (H_i - \bar{H}_i) \nablap \bar{H}_i -(H_i - \bar{H}_i)^2 \nablap \kappa_i],
\end{equation}
where we assume that $\nablap \kappa_i = \kappa_c$, and $\nablap \bar{H}_i = \bar{H}_c$ where $\kappa_c$ and $\bar{H}_c$ are constant parameters defined in \cref{sec: robust endocytic budding mechanism}.
The first term of \cref{eqn: bending chemical potential} is the shape mismatch penalty; considering the binding of a rigid protein which induces a significant spontaneous curvature change, if the curvature of membrane is far from this new spontaneous curvature, then the shape mismatch between the membrane and proteins will prevent binding.
Alternatively, if the protein is more flexible, a shape mismatch results in a small energetic penalty. 
The second term of \cref{eqn: bending chemical potential} endows the protein with curvature sensitive binding.

The in-plane diffusion of the protein is accounted for by the chemical derivative of the smoothing Dirichlet energy, $E_d$, 
\begin{equation}
    \mu^d_{i} = -\nablap E_d = -\int \eta \Delta_s \phi_i,
\end{equation}
where $\eta$ is the same Dirichlet energy constant introduced in \cref{eqn: dirichlet energy}. 
The total chemical potential captures the bending, adsorption and diffusion components.
A mobility rate constant, $B$, determines the time scale of the chemical response,
\begin{equation}
    \dot{\Bphi} = B \Bmu = B(\Bmu_b + \Bmu_{a} + \Bmu_{d}).
\end{equation}

We investigate the influence of curvature dependent binding to a realistic dendritic spine geometry which was reconstructed from electron micrographs and curated using \texttt{GAMer 2} (\cref{fig: spine}A)~\cite{lee20203d}.
The mean curvature of the spine geometry is shown \cref{fig: spine}C.
We isolate the effect of curvature-dependent binding by assuming that the shape of the spine is fixed and impose Dirichlet boundary conditions at the base on the spine to fix the protein concentration, $\phi = 0.1$ (\cref{fig: spine}A).

Starting from a homogeneous protein distribution, $\phi_{0} = 0.1$, \cref{fig: spine}B and \cref{mov: spine} show the evolution of the protein distribution and a trajectory of the system energy.
Note that for simplicity, we have turned off the adsorption energy term since it only shifts the basal protein-membrane interactions which will also be set by the Dirichlet boundary condition.
\mem3dg constrains the range of $\phi \in (0,1)$ using the interior point method~\cite{nocedalNumericalOptimization2006}.
Due to the curvature sensitivity of the protein, illustrated by the snapshots (\cref{fig: spine}B, $T = 350$) representing the final protein distribution, the protein aggregates towards regions of high curvature (e.g., on the spine head).

Although the proteins can reduce the bending energy by modulating the local bending modulus and spontaneous curvature, the protein distribution at equilibrium does not cancel out the bending energy.
We expect that the Dirichlet energy term, which limits $\phi$ to be smooth across the geometry, restricts the protein from further aggregating to the extent required to cancel out the bending energy.
The components of forces on the initial and final configurations of the spine are compared in \cref{fig: spine}D-F. 
The initial homogeneous protein distribution has no line tension forces and a bending force shown in \cref{fig: spine}D. 
After the protein distribution reaches steady state, line tension appears in response to membrane heterogeneity \cref{fig: spine}E. 
We hypothesize that, similar to \cref{sec: robust endocytic budding mechanism}, the line tension constricts the neck of the spine and helps to support the cup-like structures in the spine head.
While in most regions the distribution of proteins reduces the force, several regions experience increased stress \cref{fig: spine}F.
Note that the magnitude of the forces generated by proteins in this model are orders of magnitude smaller than the initial bending force.
Thus, this example demonstrates that \mem3dg can be used on meshes imported from realistic geometries of cellular substructures. 

\subsection{Membrane dynamics with full mechanochemical feedback}\label{sec: protein variation}

\begin{table}[htb]
    \caption{Parameters used in \cref{sec: protein variation} for models with full mechanochemical feedback} \label{table: parameter in bud on vesicle}
    \begin{center}
        \begin{tabular}
            {@{}cc@{}}
            Parameters  &Values   \\ \toprule
            $\phi_0 $ & 0.1 \\
            $\kappa_c $ & \SI{0}{\nano \newton \um} \\
            $\bar{H}_c $ & \SI{10}{\per \um} \\
            $K_V$  & \SI{0.5}{\nano \newton \um}           \\
            $K_A$  & \SI{1}{\nano \newton \per \um}    \\
            $B$     & \SI{3}{\per \nano \newton  \per \um  \per \second }  \\
            $\varepsilon$  & \SI{-1e-3}{\nano \newton \um }  \\
            $\eta$  & \SI{0.1}{\um \nano \newton}   \\    
            $\bar{V}$ & 2.91, 3.95, 4.99 \si{\um \cubed} \\
        \end{tabular}
    \end{center}
\end{table}

\begin{figure}[htbp]
    \centering
    \includegraphics[width=\linewidth]{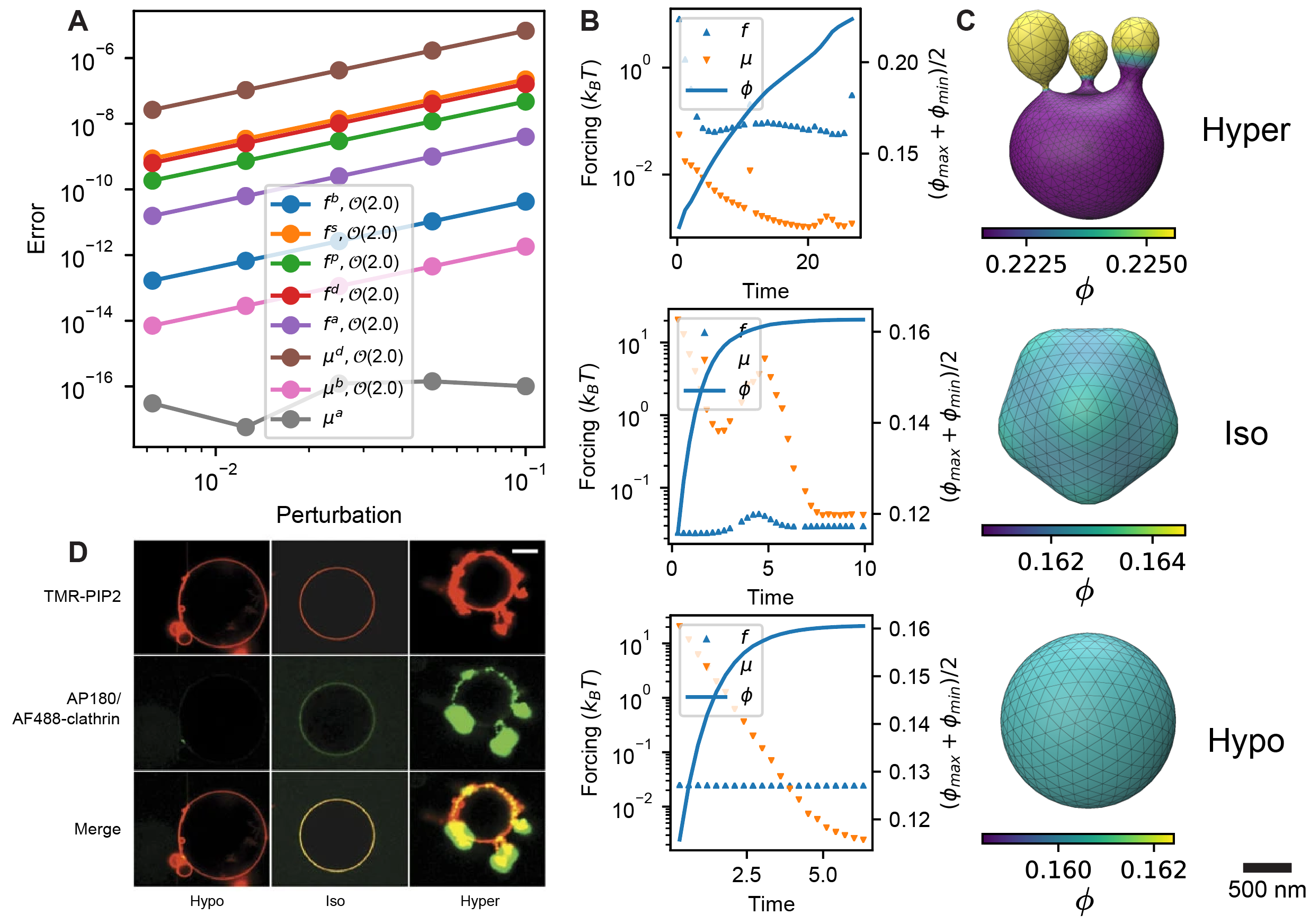}
    \caption{
        Modeling budding from a vesicle driven by the full mechanochemical feedback of membrane--protein interactions. 
        A) Validation of the exactness of the discrete forcing with respect to the discrete energy. 
        The terms correspond to forces from bending $f_b$, tension-area $f_s$, pressure-volume $f_p$, Dirichlet $f_d$, and protein binding $f_a$.
        $\mu_d$, $\mu_b$, and $\mu_a$ are the chemical potential of diffusion, bending, and binding respectively. 
        B) The time trajectory of budding dynamics in hypertonic, isotonic, and hypotonic osmotic condition. 
        C) The final snapshot of the system configuration under hypertonic, isotonic, hypotonic condition. 
        D) Similar geometries to those shown in C) have been observed in experiments by \textcite{saleemBalanceMembraneElasticity2015}.
    }
    \label{fig: bud on vesicle}
\end{figure}
\begin{movie}
    \caption{Animated time series simulation of the hypertonic case shown in \cref{fig: bud on vesicle}B-C. The color map shows the order parameter for protein density, $\phi \in (0,1)$, and $t$ represents the numerical time. Available on GitHub: \url{https://github.com/RangamaniLabUCSD/2020-Mem3DG-Applications/blob/master/examples/videos/compressed/bud_hyper_1.mp4}}
    \label{mov: bud_hyper}
\end{movie}
\begin{movie}
    \caption{Animated time series simulation of the isotonic case shown in \cref{fig: bud on vesicle}B-C. The color map shows the order parameter for protein density, $\phi \in (0,1)$, and $t$ represents the numerical time. Available on GitHub: \url{https://github.com/RangamaniLabUCSD/2020-Mem3DG-Applications/blob/master/examples/videos/compressed/bud_iso_1.mp4}}
    \label{mov: bud_iso}
\end{movie}

In this section, we will demonstrate the use of \mem3dg to model the complete mechanochemical feedback of a protein-membrane system.
For the following simulations, not only can proteins bind in a curvature-dependent manner, the membrane can also deform leading to a feedback loop.
We have introduced all of the force terms in previous sections except the shape derivative of the adsorption energy,
\begin{equation}
     \int \vf^a_i = - \nablar E_a =  - 2 \varepsilon \phi_i \int \vH_{i},
\end{equation}
which accounts for the change in the area of protein coverage (i.e., expanded coverage if $\varepsilon < 0$). 

Revisiting the claim that all discrete forcing is exact with respect to the discrete energy, we validate by examining the convergence of the forcing terms with respect to the size of perturbation to the system configuration, $\epsilon$ (\cref{fig: bud on vesicle}A). 
This is based on the leading order expansion done in \cref{eqn: taylor expansion}, which concludes that the forcing terms are exact if their rate of convergence is at least \nth{2} order. 
Shown in \cref{fig: bud on vesicle}A, this is true for all forcing terms; note that since the adsorption energy, $E_a$, is a linear function with respect to $\phi$, $\mu^a$ can be determined to the machine precision for all perturbation sizes.
A meaningful discrete-smooth comparison of all terms in the energy and forcing similar to \cref{sec: practical considerations for applying Mem3DG to biological problems} requires the analytical solutions of the bending force and interfacial line tension arising from the spatially heterogeneous membrane properties, which to the best of our knowledge, are not available.
In \cref{sec: robust endocytic budding mechanism}, we introduced a heterogeneous membrane composition as a static property. 
By prescribing the protein density profile, we are able to get hints about how to form membrane buds from a patch. 
Here we lift this assumption and simulate the dynamics of osmotic pressure-driven budding from a vesicle. 
The dynamics couples the protein-membrane mechanochemical feedback and includes protein binding and diffusion introduced in \cref{sec: spine}.
The expressions of discrete free energy and forcings do not change but we allow the membrane configuration and protein density to interact and evolve simultaneously.

We start each simulation from a sphere with a uniform protein concentration, $\varphi= \varphi_0 = 0.1$.
We consider the evolution of the system in different osmotic conditions: hyper-, iso-, and hypotonic. 
Additional parameters for these simulations are given in \cref{table: parameter in bud on vesicle}.
\cref{fig: bud on vesicle}B shows plots of the mechanical, $\|\vbf\|_{L_2}^2$, and chemical response, $\|\Bmu\|_{L_2}^2$, along with the protein density, $(\phi_{\text{max}} + \phi_{\text{min}})/2$, over the trajectory for each osmotic condition.
Note that under hypo- and isotonic conditions, the system reaches (quasi) steady state where both the shape and protein distribution stabilize, while in hypertonic solution, the system continues to experience strong mechanical force and protein mobility, which we expect to drive further morphological changes of the membrane beyond our simulation stopping point. 
\cref{fig: bud on vesicle}C shows the final snapshot of each simulation across the osmotic conditions with the protein density represented by the color map.
In hypertonic conditions, the osmotic pressure provides sufficient perturbations to membrane morphology, which initializes the positive feedback loop between membrane curvature generation and protein aggregation;
This mechanochemical feedback jointly promotes the outward bending of the membrane and results in the bud formation (\cref{fig: bud on vesicle}C--hyper and \cref{mov: bud_hyper}).
Under isotonic and hypotonic conditions, the osmolarity does not permit the large change in volume required to form spherical buds with small neck radius.
In hypotonic condition, the pressure-tension balance provides substantial stability to the initial spherical configuration.
In comparison, in the isotonic condition, the comparable competition between the chemical and mechanical response leads to a patterned protein distribution and an undulating morphology (\cref{fig: bud on vesicle}C--hypo and \cref{mov: bud_iso}).
This example illustrates the possibility to utilize \mem3dg to model a full mechanochemical feedback between membrane and protein.
Although we do not intend to replicate the exact experimental conditions and assumptions, the geometries obtained from these simulations resemble the shapes obtained by \textcite{saleemBalanceMembraneElasticity2015} who investigated budding from spherical vesicles under differing osmotic conditions (\cref{fig: bud on vesicle}D)~\cite{saleemBalanceMembraneElasticity2015}.

\section{Summary}
In this work, we introduce a new perspective for constructing a 3D membrane mechanics model on discrete meshes.
The goal of our approach is to close the gap between existing discrete mesh based models~\cite{gompper1996random, julicher1996morphology,kantor1987phase,bianBendingModelsLipid2020,brakkeSurfaceEvolver1992,jieNumericalObservationNonaxisymmetric1998,krollConformationFluidMembranes,atilganShapeTransitionsLipid2007a,bahramiFormationStabilityLipid2017,noguchiShapeTransitionsFluid2005a, tachikawaGolgiApparatusSelforganizes2017,tsaiRoleCombinedCell2020,pezeshkian2019multi,sadeghiParticlebasedMembraneModel2018} and the smooth theory.
Specifically, we seek to advance the discussion behind the choice of algorithmic approaches for computing geometric values required for the discrete energy and force ~\cite{bianBendingModelsLipid2020,guckenbergerBendingAlgorithmsSoft2016,guckenbergerTheoryAlgorithmsCompute2017}.
We start by writing a discrete energy, \cref{eqn: discrete free energy}, mirrorring the spontaneous curvature model.
Then using the perspective of \gls{ddg}, we show that there is a formulaic approach for deriving the corresponding discrete force terms based on fundamental geometric vectors.
By identifying geometric invariants and grouping terms, the resulting discrete forces have exact correspondence to the traditional smooth theory.
This helps us to facilitate the comparison between smooth and discrete contexts enabling new geometric perspectives and understanding of numerical accuracy.
Moreover, the discrete force terms are functions of readily accessible geometric primitives, and the formulation is amenable for efficient implementation and extension.

We have produced a reference software implementation called \mem3dg.
Using \mem3dg, we validate our theory by reproducing the solutions to the classical shape transformations of a spherical and tubular vesicle.
We further demonstrate the derivation and incorporation of additional physics terms to model protein-membrane interactions.
Following our formulaic approach using \gls{ddg}, we derived the discrete analog of various physics such as the interfacial line tension, surface-bulk adsorption, protein lateral diffusion, and curvature-dependent protein aggregation. 
To exemplify all the introduced physics, the full mechanochemical coupling between the membrane shape and protein density evolution results in protein localization, pattern formation, and budding.
These examples serve to highlight the extensibility of the framework, which does not require the introduction of coordinates and advanced tensor calculus commonly needed to solve problems on arbitrary manifolds. 
The software implementation \mem3dg was designed to facilitate coordination between theoretical modeling and wet-experiments; and we hope to support the modeling of scenes with well resolved protein-membrane interactions such as in the electron tomograms \cite{SerwasEtAl2021}. 
We expect that as the advances in biophysical modeling and membrane ultrastructure imaging progresses, \mem3dg will emerge as a useful tool to test new hypothesis and understand cellular mechanobiology.

\section{Acknowledgments}

The authors would like to acknowledge Dr.~Ali~Behzadan, Prof.~Ravi~Ramamoorthi, and Prof.~Albert~Chern for helpful discussions and critical feedback.
This work was supported in part by the National Institutes of Health R01GM132106, National Science Foundation DMS 1934411, and an Air Force Office of Scientific Research DURIP FA9550-19-1-0181 to PR.
CTL also acknowledges support from a Hartwell Foundation Postdoctoral Fellowship.

\clearpage

\appendix
\appendixpage
\counterwithin{figure}{section}
\counterwithin{table}{section}

\section{Supplemental figures}

\begin{figure}[htbp!]
    \centering
    \includegraphics[width=5in]{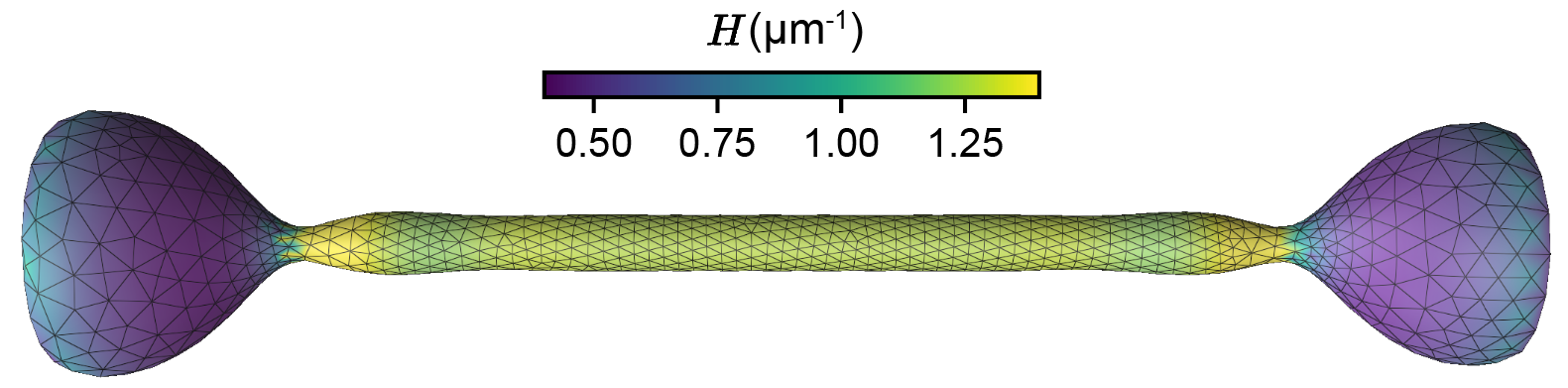}
    \caption{
        A symmetric metastable state with two beads instead of single larger bead is observed, prior to collapsing to the solution shown in \cref{fig: homogeneous}C -- high tension
    }
\end{figure}
\label{SI_fig: metastable two beads}

\begin{figure}[htbp!]
    \centering
    \includegraphics[width=\linewidth]{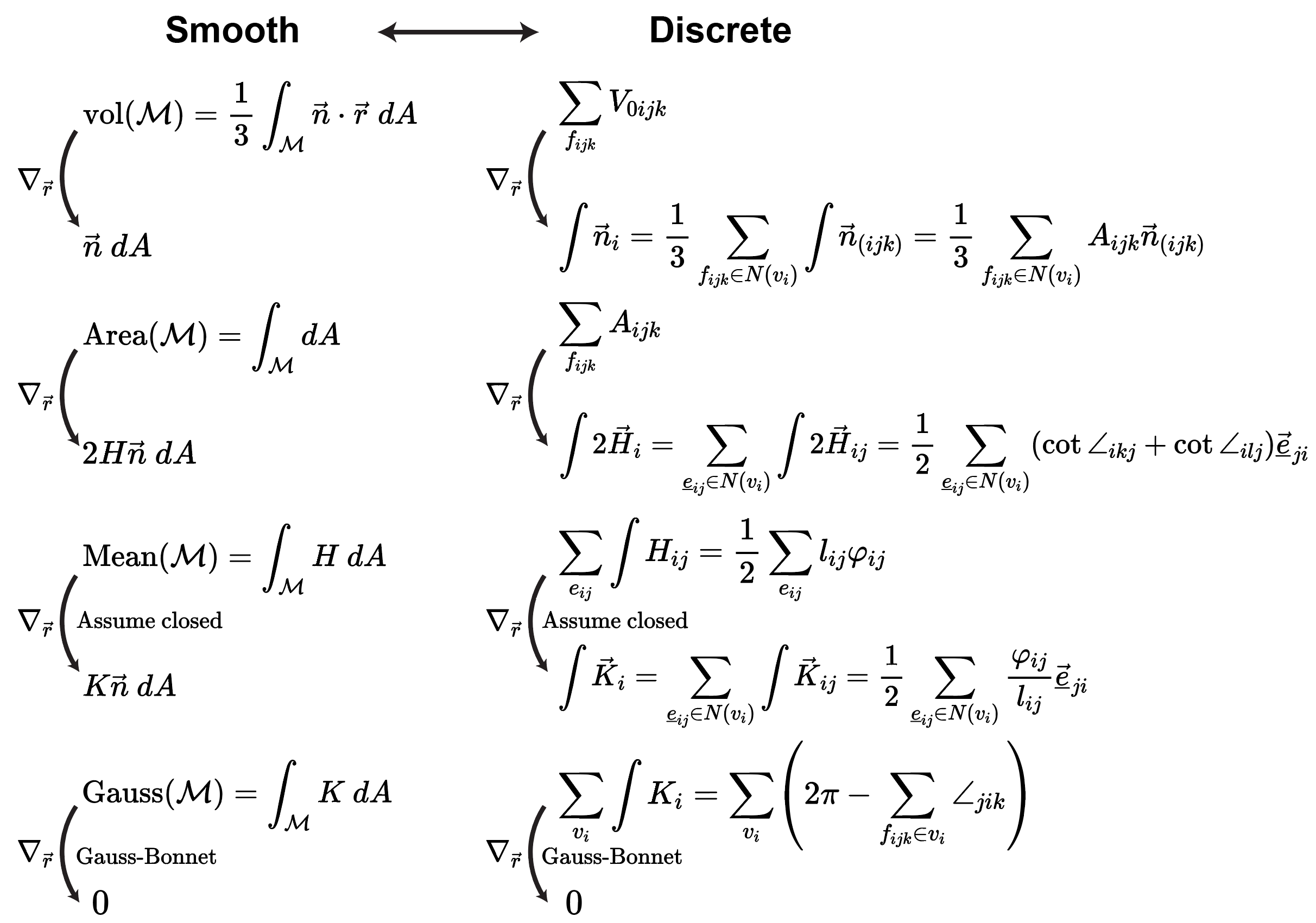}
    \caption{
        Steiner's formula in continuous and discrete geometry: chain of smooth and discrete shape derivatives of integrated geometric measurements~\cite{craneLectureSlide}.
        }
    \label{SI_fig: chain of variation}
\end{figure}

\section{Rationale for integrated measurements in discrete contexts}
\label{sec:integrated_measures}

The rationale for why an integrated measurement in discrete contexts is the natural counterpart to pointwise measurements in smooth contexts can be demonstrated by considering the curvature of a discrete polygonal curve. 
If we attempt to define the curvature, $C$, of the discrete polygonal curve in a na\"ive pointwise manner, following the procedure in smooth settings, we will find zero curvature along edges and infinite curvature (owing to the discontinuity) on vertices.
Thus the traditional view of curvature from smooth manifolds reveals no useful information about the geometry of the discrete curve.
We must find another geometric relationships which can translate between smooth and discrete contexts to maintain the geometric connection.

One relationship from smooth differential geometry is the equivalence of the integrated curvature and the turning angle $\psi$ (i.e., the total angle by which the tangent vector of the curve turns over some domain $l$).
Returning to the discrete context, we can seek to preserve this relationship between the integrated curvature and turning angle by finding a compatible definition.
Since the discrete turning angle, $\psi_i$, between two connected edges of the discrete polygonal curve is well-defined, we can set the ``discrete'' curvature, $\int C$, of a vertex, $v_i$, to be 
\begin{equation}
    \left(\int C\right)_i \equiv \psi_i.
\end{equation}
\emph{We note that the notation for the discrete curvature, $\left(\int C\right)_i$ is used only in this illustrative example; in the remainder of the text, we will omit the parenthesis and use the simplified notation, $\int C_i$.}
To make sense of the integral over a discrete object, additional care must be taken to represent the curvature from a distributional sense~\cite{chernDDG}.
This is related to traditional approximation methods, such as the point allocation method, which bridges a smooth and discrete problem by convoluting the smooth problem with impulse functions (e.g., the Dirac delta function) at a finite number of locations~\cite{hughes2012finite}.

As we have shown, integrated geometric measurements enable us to preserve geometric relationships (from smooth contexts) for discrete objects, and are thus preferred over pointwise definitions.
Nevertheless, we often require a pointwise discrete measurement for use in algorithms and visualization. 
An integrated measurement can be converted to a meaningful pointwise discrete measurement by normalizing the value over a domain.
For the discrete polygonal curve, the domain can be the dual vertex length, $l_i$ (i.e., the discrete analog of $l$).
$l_i$ is given by half of the sum lengths of the two incident edges.
A pointwise curvature on the vertex $v_i$ is then given by, 
\begin{equation}
    C_i = \int C_i/l_i = \psi_i/l_i. 
\end{equation}

Another rationale for using an integrated value for a discrete geometric measurement, is that we can arrive at the same definition from multiple perspectives.
Returning to the definition of the curvature of a polygonal curve, without introducing the turning angle, we can arrive at the same result by adopting the Steiner view ~\cite{chernDDG, steiner_2013}\footnote{We will use the Steiner view to define the discrete curvature of a surface in \cref{sec: discrete energy defined by mesh primitives}}. 
In the Steiner view, we replace the sharp vertices with a smooth circular arc with radius $\epsilon$ such that the discrete geometry is made smooth such that the curvature is well-defined everywhere. 
As the only curved section, every circular arc has a discrete (integrated) curvature,
\begin{equation}
    \int C = \int_{\mathrm{arc}} C\,ds = C_{\mathrm{arc}} l_{\mathrm{arc}} = \frac{1}{\epsilon} (\epsilon \psi) = \psi,
\end{equation}
where $C_{\mathrm{arc}} = 1/\epsilon$ is the curvature of the circular arc, and $l_{\mathrm{arc}} = \epsilon\psi$ is the arc length.
We see that in the Steiner view, the integrated curvature is still equivalent to the turning angle. 
Following similar logic, other discrete definitions are described in \cref{sec: discrete energy defined by mesh primitives} and the \gls{ddg} literature ~\cite{chernDDG,craneDISCRETEDIFFERENTIALGEOMETRY}.

\section{Discrete shape and chemical derivatives of discrete energy}
\subsection{Halfedge on a triangulated mesh}\label{sec: halfedge}
\begin{figure}[htbp]
    \centering
    \includegraphics[width=4 in]{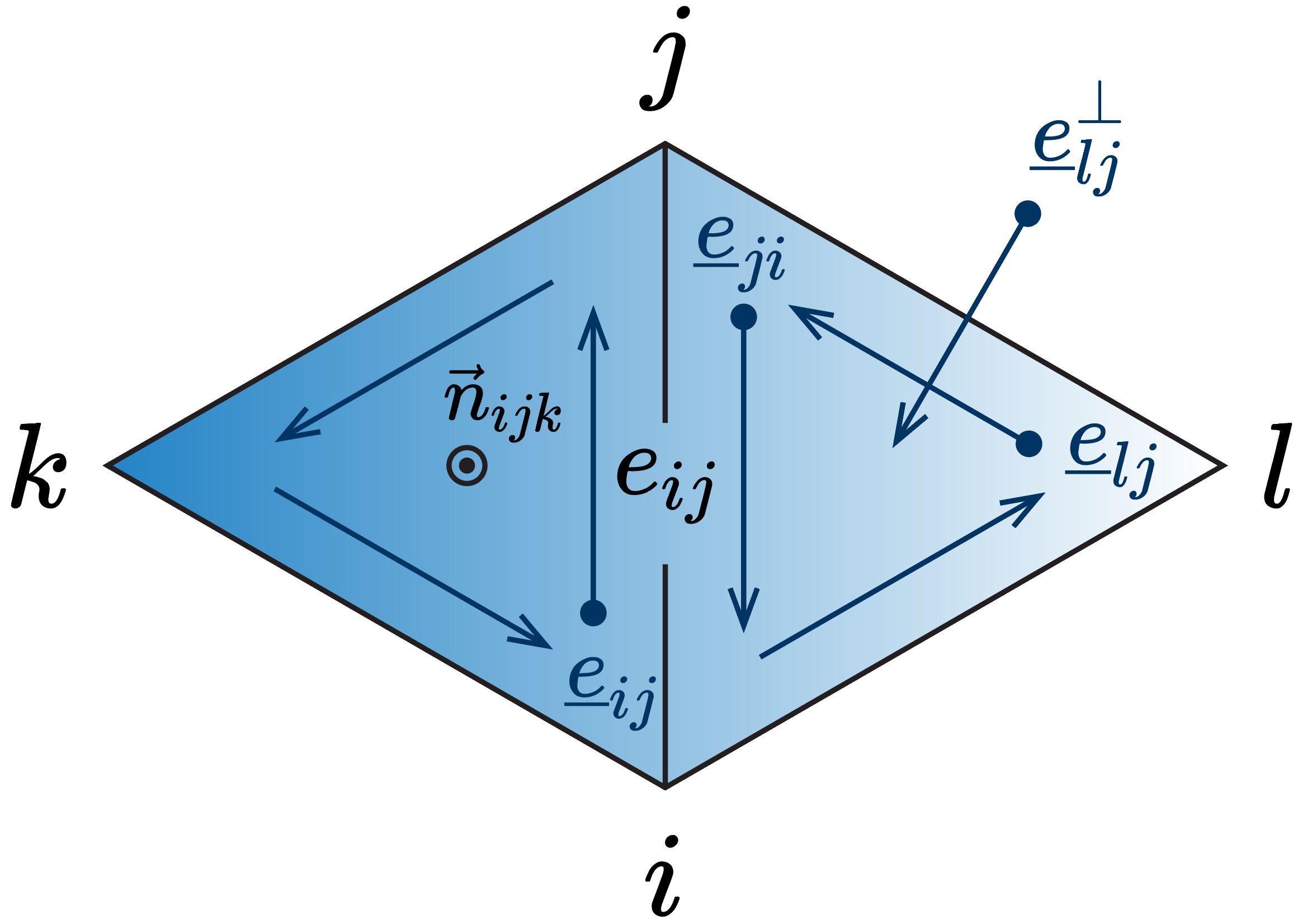}
    \caption{Schematics for halfedge on a triangulated mesh. }
    \label{SI_fig: halfedge schematics}
\end{figure}
A scalar quantity on an edge is symmetric with respect to index permutation;
For example, the scalar mean curvature (\cref{eqn: scalar mean curvature}),
\begin{equation}
    \int H_{ij} = \int H_{ji} = \frac{l_{ij} \varphi_{ij}}{2}.
\end{equation}
However, as we will show in detail in the following sections, this symmetry does not apply to vector quantities, which compose the discrete shape derivative of the energy, force. 
For example, the corresponding mean curvature vector, 
\begin{equation}\label{eqn: asymmetry with vector measure}
    \int \vH_{ij} \neq \int \vH_{ji}.
\end{equation}
To highlight the directionality of vector quantities and disambiguate the notation, here we review the concept of halfedge on a triangulated mesh. 
Given any non-boundary edge, $e_{ij}$, on a manifold mesh, there exits two associated halfedges, $\underline{e}_{ij}$ and $\underline{e}_{ji}$ (\cref{SI_fig: halfedge schematics}).
This convention leads to an oriented (counterclockwise) halfedge loop on each triangle face and subsequently a well-defined 1) \ang{90} counterclockwise rotation of the halfedge in the plane of the face (e.g., $\underline{e}_{lj} \rightarrow \underline{e}_{lj}^{\perp}$), and 2) face normal (outward) based on the right hand rule \cref{SI_fig: halfedge schematics}. 
Beside being used to differentiate vector/scalar quantities, the concept of halfedge is widely adopted data structure for managing connected graph, or meshes, for which we refer the reader to the broader literature~\cite{geometrycentral,libigl}. 

\subsection{Deriving the bending force as the shape derivative of bending energy} 
\label{sup_sec: derive bending force}

The geometric derivatives of mesh primitives, including edge length, $l$, dihedral angle, $\varphi$, vertex dual area, $A$, are given as 
\begin{subequations}\label{eqn: basic geometric derivatives}
    \begin{align} 
        &\nablari l_{ij} = \frac{\vec{\underline{e}}_{ji}}{l_{ij}}, \label{eqn: edge length variation} \\
        &\nablari \varphi_{ij} = \frac{1}{l_{ij}} (\cot\angle_{ijk}\vn_{ijk} + \cot\angle_{ijl}\vn_{ilj}), \label{eqn: diagonal dihedral variation}\\
        &\nablari \varphi_{jk} = -\frac{1}{l_{jk}} (\cot \angle_{ijk} + \cot \angle_{ikj}) \vn_{ijk} = -\frac{l_{jk}}{2 A_{ijk}} \vn_{ijk} ,\label{eqn: offdiagonal dihedral variation}\\
        &\nablari A_i = \frac{1}{3}\nearbyFaces  \nablari A_{ijk} = \frac{1}{6} \nearbyHalfedges (\cot\angle_{ikj} +\cot\angle_{ilj})\vec{\underline{e}}_{ji},\label{eqn: diagonal area variation}\\
        &\nablari A_j = \frac{1}{3} \sum_{f_{ijk} \in N(e_{ij})} \nablari A_{ijk} = \frac{1}{6} (\vec{\underline{e}}_{jk}^\perp + \vec{\underline{e}}_{lj}^\perp) ,\label{eqn: offdiagonal area variation}
    \end{align}
\end{subequations}
where $\vn_{ijk}$ is the unit normal vector of the face $f_{ijk}$, $\vec{\underline{e}}_{ji}$ is the vector aligned with the halfedge, $\underline{e}_{ji}$, with its length of $l_{ij}$~\cite{grinspun2003discrete}. 
The indices and nomenclature in \cref{eqn: diagonal dihedral variation}, \cref{eqn: offdiagonal dihedral variation} and \cref{eqn: offdiagonal area variation} are illustrated in the diamond neighborhood (\cref{SI_fig: halfedge schematics}) and those of \cref{eqn: diagonal area variation} are illustrated in the fan neighborhood (\cref{fig: energy and force}A).

To simplify the expression and provide more structure for the subsequent discrete variation, it is convenient to define some fundamental curvature vectors, 
\begin{subequations} \label{eqn: curvature vectors}
    \begin{align} 
        &\int 2\vH_{ij} = \frac{1}{2} (\nablari A_{ijk} + \nablari A_{ijl}) = \frac{1}{4}(\vec{\underline{e}}_{jk}^\perp + \vec{\underline{e}}_{lj}^\perp)  \\
        &\int \vK_{ij} = \frac{1}{2}  \varphi_{ij}  \nablari l_{ij} \\
        &\int \vS_{ij,1} = \frac{1}{2} l_{ij} \nablari \varphi_{ij} = \frac{1}{2}(\cot\angle_{ijk}\vn_{ijk} + \cot\angle_{ijl}\vn_{ilj}) \\
        &\int \vS_{ij,2} =  \frac{1}{2} \left( l_{jk} \nablari \varphi_{jk} + l_{jl}\nablari \varphi_{jl}  + l_{ji} \nablari \varphi_{ji} \right) = -\frac{1}{2}(\cot\angle_{jki}\vn_{ijk} + \cot\angle_{ilj} \vn_{ilj}),
    \end{align}
\end{subequations}
where the mean curvature vector, $\int \vH$, results from area gradient; Gaussian curvature vector, $\int \vK$, and the Schlafli vector, $\int \vS$, consists of the two components of the variation of total mean curvature, $\frac{1}{2} \sum_{e_{ij}} l_{ij} \varphi_{ij}$. 
The asymmetry of vector quantities in \cref{eqn: curvature vectors} under index permutation (\cref{eqn: asymmetry with vector measure}) arises from the vertex we take the shape derivative with respect to (i.e., $v_i$, or $v_j$); because of the asymmetry, we can associate each Schlafli vector with a unique halfedge.
Similar to the translation from edge values to vertex value (\cref{eqn: edge value to vertex value}), we can also translate the halfedge value to vertex value by summing all halfedge values over the fan neighborhood,
\begin{equation}
    \int (\cdot)_i = \nearbyHalfedges \int (\cdot)_{ij}.
\end{equation}
Note that unlike translating edge values, there is no prefactor $1/2$ for translating halfedge values because each halfedge is uniquely associated with one vertex.
The translated curvature vectors on a vertex cane compared against vertexwise smooth analytical solutions as benchmarked in \cref{sec: practical considerations for applying Mem3DG to biological problems}. 
Now we have all of the elements needed to derive the derivatives of the discrete Willmore bending energy.
Because the discrete energy is locally supported by the vertex, $v_i$, and its 1-ring neighbors, $v_j \in N(v_i)$, we can separate them into the ``diagonal'' term, and ``off-diagonal'' term, 
\begin{align}
    \begin{split}
        \int \vf^b_i = -\nablari E_b &= - \nablari \left (\sum_i \kappa_i (H_i(\vr) - \bar{H}_i)^2 A_i(\vr) \right )\\
        &= - \underbrace{\nablari \left[ \kappa_i (H_i - \bar{H}_i)^2 A_i \right]}_{\text{diagonal}}  
        - \underbrace{\nearbyVertices \nablari \left[\kappa_j (H_j - \bar{H}_j)^2 A_j  \right]}_{\text{off-diagonal}}.
    \end{split}
\end{align}
Using the derivatives of geometric primitives in \cref{eqn: basic geometric derivatives}, we can assemble the derivatives of local pointwise mean curvature for both the ``diagonal'' term,  
\begin{align}
    \begin{split}
        \nablari H_i &= \frac{1}{4} \nearbyHalfedges  \nablari \frac{ l_{ij} \varphi_{ij}}{A_i} \\
        &= \frac{1}{4A_i} \nearbyHalfedges \left (\varphi_{ij} \nablari l_{ij} + l_{ij} \nablari \varphi_{ij} \right) - \frac{H_i}{A_i} \nablari A_i \\
        &= \frac{1}{A_i} \nearbyHalfedges \frac{1}{2}\left (\int \vK_{ij} + \int \vS_{ij,1} \right) - \frac{2}{3} H_i \int \vH_{ij} ,
    \end{split}
\end{align}
and for the ``off-diagonal'' term,
\begin{align}
    \begin{split}
        \nablari H_j &= \frac{1}{4} \sum_{\underline{e}_{jk} \in N(v_j)}  \nablari \frac{ l_{jk} \varphi_{jk}}{A_j} \\
        &= \frac{1}{4A_j}  \left ( l_{jk} \nablari \varphi_{jk} + l_{jl}\nablari \varphi_{jl}  + \varphi_{ji}\nablari l_{ji}  + l_{ji} \nablari \varphi_{ji} \right) - \frac{H_j}{A_j} \nablari A_j\\
        &= \frac{1}{2A_j}\left(\int \vK_{ij} + \int \vS_{ij,2}\right ) - \frac{4}{3} H_j \int \vH_{ij} .
    \end{split}
\end{align}
When written in the halfedge form, factoring out the fundamental curvature vectors introduced in \cref{eqn: curvature vectors}, we obtain the discrete bending force as 
\begin{align}
    \begin{split}
        \int \vf^b_i = \nearbyHalfedges & -\left[ \kappa_i (H_i - \bar{H}_i) + \kappa_j (H_j - \bar{H}_j) \right] \int \vK_{ij}\\
        + &\left[\frac{1}{3} \kappa_i (H_i - \bar{H}_i) (H_i + \bar{H}_i) + \frac{2}{3} \kappa_j (H_j - \bar{H}_j) (H_j + \bar{H}_j) \right] \int 2 \vH_{ij} \\
        - &\left[ \kappa_i (H_i - \bar{H}_i) \int \vS_{ij,1} + \kappa_j (H_j - \bar{H}_j) \int \vS_{ij,2}\right].
    \end{split}
\end{align}

\subsection{Deriving the line tension and diffusion as the shape and chemical derivatives of the Dirichlet energy} 
\label{sup_sec: derive line tension and diffusion as variations of Dirichlet energy}

Since the discrete Dirichlet energy is constructed on the triangular face and therefore does not involve any neighborhood, we simplify the notation by adopting the convention illustrated in \cref{fig: energy and force}C.
The gradient of protein density is given by the slope of the fitted plane over the vertexwise protein density, which is piecewise constant for each face, 
\begin{align}
    \nablat ~\phi_i = \frac{1}{2 A_{ijk}}\sum_{\vec{\underline{e}}_{k} \in N(f_{ijk})} \phi_k \ve_k^{~\perp},
\end{align}
where we adopt the counterclockwise convention (e.g. $\ve_k = \vec{\underline{e}}_{ji}$) and $(\cdot)^\perp$ represents a \ang{90} counterclockwise rotation in plane of the face, $f_{ijk}$.

\subsubsection{Line tension from the shape derivative of the Dirichlet energy}

Substituting the definition of the discrete gradient into the Dirichlet energy (\cref{eqn: dirichlet energy}), we expand the energy in terms of mesh primitives, whose geometric derivatives are given in \cref{eqn: basic geometric derivatives}.  
Additional formulae are needed to compute the geometric derivatives of the outer angles of the triangle (\cref{fig: energy and force}C)
\begin{subequations}\label{eqn: derivatives of angles}
\begin{align} 
    \nablari \angle_k &= \frac{\vn \times \ve_j}{ \| \ve_j \|^2} \\
    \nablari \angle_j &= \frac{\vn \times \ve_k}{ \| \ve_k \|^2} \\
    \nablari \angle_i &= -(\nablari \angle_k + \nablari \angle_j), 
\end{align}
\end{subequations}
which arise from the calculation of the $L_2$ norm of the gradients as the result of vector inner product.
When combined, the geometric derivatives for the quadratic gradient term is
\begin{align}
    \begin{split}
        \quad \nablari & \left\langle \sum \phi_k \ve_k^{~\perp}, \sum \phi_k \ve_k^{~\perp} \right \rangle = \\
        & \quad + \phi_k  \phi_k  \ve_k - 2 \phi_j \phi_j  \ve_j\\
               & \quad + 2  \phi_j  \phi_i  \| \ve_i \| \left(-\hat{e}_j  \cos\angle_k + \| \ve_j \|\nablari (\cos\angle_k) \right) \\
               & \quad + 2  \phi_i  \phi_k  \| \ve_i \| \left(\hat{e}_k  \cos\angle_j + \| \ve_k \|  \nablari (\cos\angle_j) \right)\\ 
               & \quad + 2  \phi_j  \phi_k  \left(-\hat{e}_j  \| \ve_k \| \cos\angle_i +
                    \| \ve_j \|  \hat{e}_k  \cos\angle_i +
                    \| \ve_j \|  \| \ve_k \|  \nablari (\cos\angle_i) \right)
    \end{split}
\end{align}
Then we can get the final shape derivative by combining the area gradient, or the mean curvature vector (\cref{eqn: curvature vectors}).  

\subsubsection{Surface diffusion from the chemical derivative of the Dirichlet energy}

In the case where we are evolving the protein distribution, we need the chemical derivative of the Dirichlet energy. 
Before we look into the discrete case, we can first tackle the problem in the smoooth setting, which is a classic textbook example. 
Using the Green's first identity, or integration by parts on a 2-manifold, 
\begin{equation}
    \int_{\mathcal{M} }(\psi \Delta_s \varphi+\nablat  \psi \cdot \nablat  \varphi) dA=\oint_{\partial \mathcal{M}} \psi \nablat  \varphi \cdot \vn d S, 
\end{equation}
and ignoring the boundary term at the right hand side, we arrive at an alternative expression for the Dirichlet energy, 
\begin{equation}
    E_d = \frac{1}{2} \int_\mathcal{M} \eta \| \nablat \phi \|^2~ dA = -\frac{1}{2} \int_\mathcal{M} \eta \phi \Delta_s \phi ~ dA.
\end{equation}
The same procedure can be followed in the discrete case.
The discrete Dirichlet energy (\cref{eqn: dirichlet energy}) can be written in matrix form, 
\begin{equation}
    E_d = \frac{1}{2} \eta \Bphi^\top \tilde{\bG}^\top  \tilde{\bT} \tilde{\bG} \Bphi
\end{equation}
where $\tilde{\bG}$ is the gradient tensor which maps scalar value on vertices to vector values on faces, and $\tilde{\bT} = \mathrm{diag} (\bA^{\text{face}})$ is the $|f| \times |f|$ diagonal matrix with entries corresponding to the area of each mesh triangle face. 
Through integration by parts on a discrete geometry, the discrete Dirichlet energy can be equivalently expressed as
\begin{equation}
    E_d =  \frac{1}{2} \eta \Bphi^\top \tilde{\bL} \Bphi, 
\end{equation}
which is a quadratic form with respect to the cotangent Laplacian matrix, $\tilde{\bL}$ ~\cite{chernDDG, craneDISCRETEDIFFERENTIALGEOMETRY}. 
The chemical derivative of the Dirichlet energy, or the diffusion potential, is
\begin{equation}
    \Bmu^d = -\nablap E_d = -\eta \int \Delta_s \Bphi = -\eta \tilde{\bL} \Bphi.
\end{equation}
In other words, the chemical gradient flow of the Dirichlet energy is the diffusion equation. 
Note that $\tilde{\bL} =  \tilde{\bG}^\top  \tilde{\bT} \tilde{\bG}$, $\tilde{\bG}^\top$ is referred to as the discrete divergence operator that maps face vectors to scalars on vertices~\cite{libigl}.

\section{Discrete-smooth comparison on spheroid}\label{sec: spheroid benchmark}
The smooth-discrete comparison is done on the spheroid with the parametrization, 
\begin{align}
    (x, y, z) = \left ( a \cos \beta \cos \theta, a \cos\beta  \sin\theta , c \sin\beta  \right),
\end{align}
where $a = 1$, $b = 0.5$, $\beta$ is the parametric latitude and $\theta$ is the azimuth coordinate.
All geometric measurements of the smooth geometry used for benchmarking were obtained using the symbolic algbra software \textit{Sympy}. 
The corresponding discrete measurements are computed using \mem3dg, whose input spheroid mesh is mapped from a subdivided icosphere. 
The subsequent error norms for local measurements are computed based on definitions used in \cref{sec: defining metrics for convergence and comparison of forces}.

\section{Mesh regularization and mesh mutation}\label{sup_sec: mesh regularization and mesh mutation}

\begin{figure}[htbp!]
    \centering
    \includegraphics[width=\linewidth]{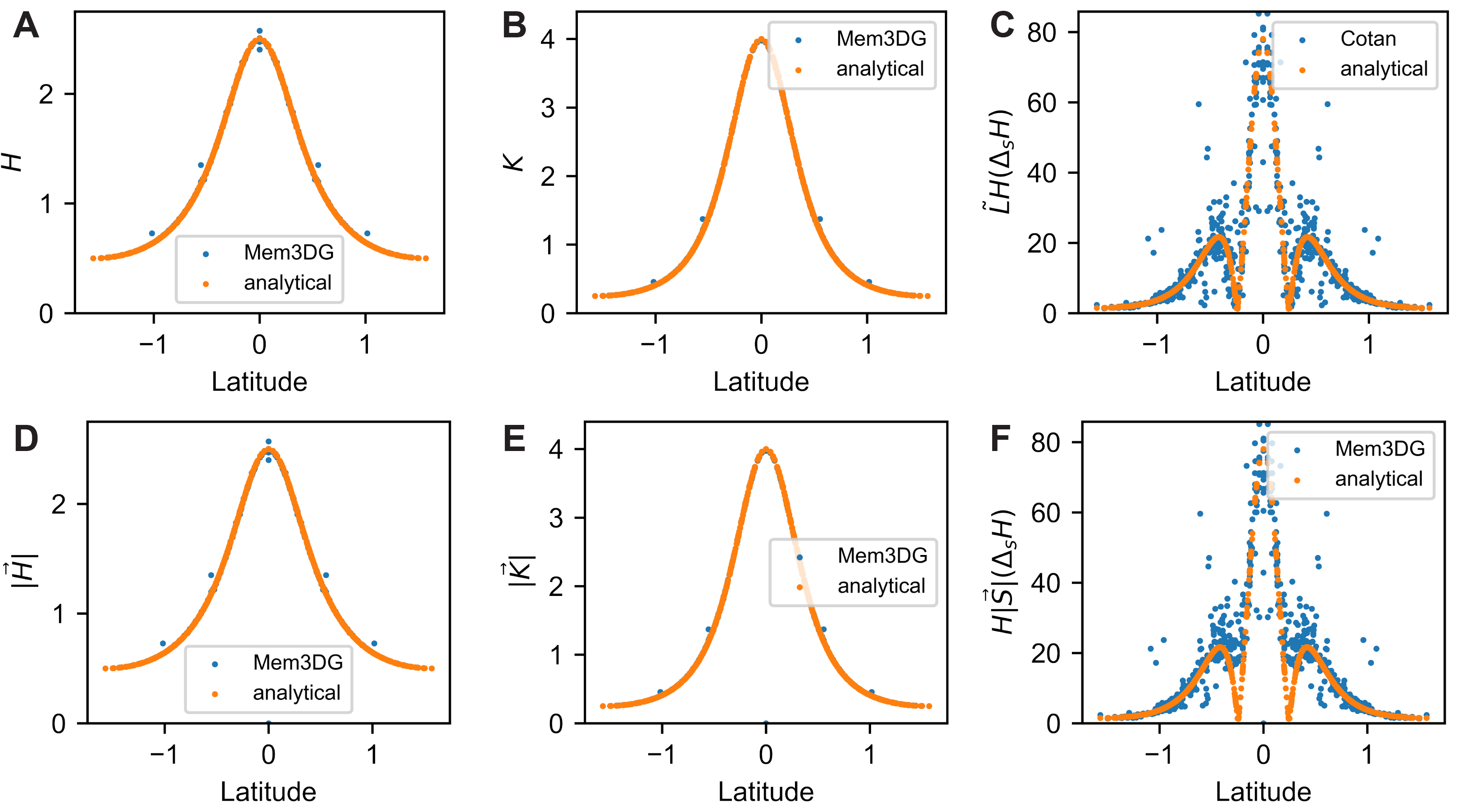}
    \caption{
        Pointwise magnitude comparison of continuous and discrete measurements: A) scalar mean curvature, B) scalar Gaussian curvature, C) (scalar) bi-Laplacian term $\nabla H$ based on cotan formula, D) vector mean curvature, E) vector Gaussian curvature, and F) (vector) bi-Laplacian term based on Schlafli vector. 
        Note that the result of the cotangent Laplacian approach in C) produces a scalar result while our approach using the Schlafli vector in F) is a vector result, thus their direct comparison is not meaningful.
    }
    \label{SI_fig: ptwise magnitude}
\end{figure}

\begin{figure}[htbp!]
    \centering
    \includegraphics[width= 6 in]{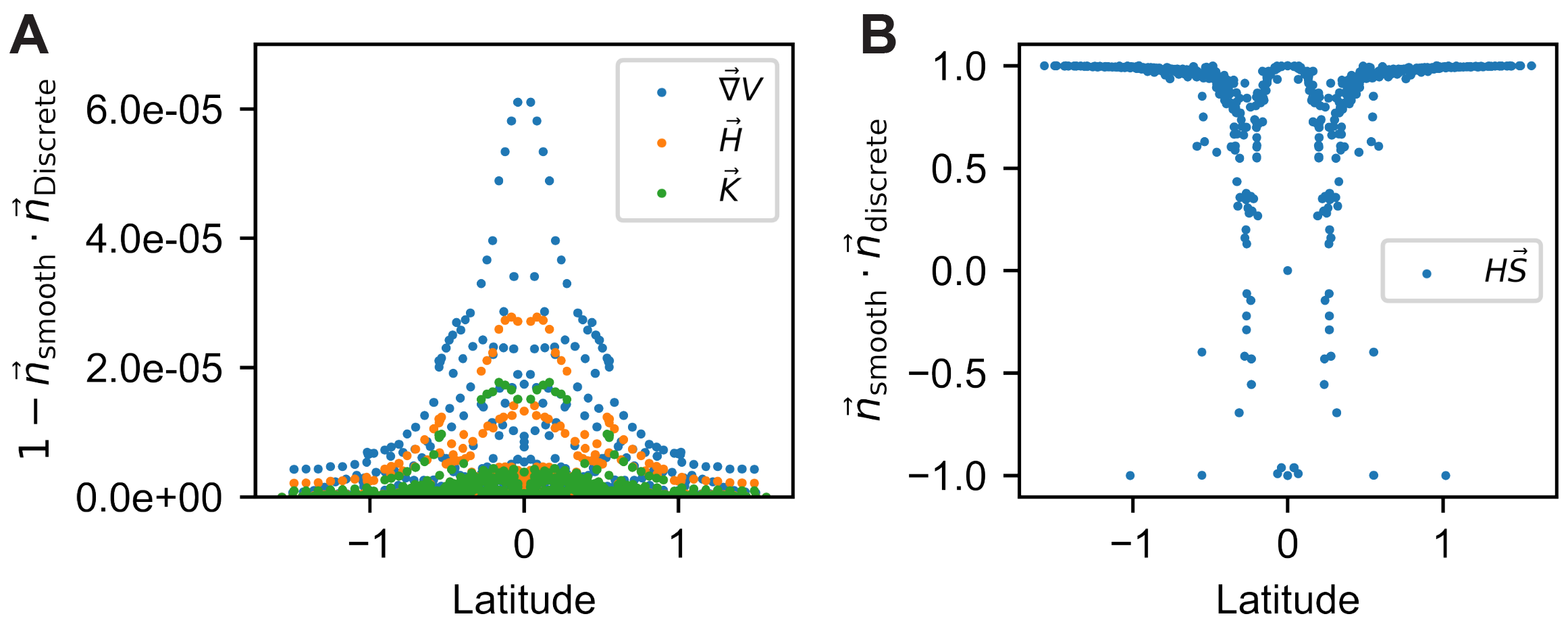}
    \caption{Pointwise directional comparison of continuous and discrete measurements: discrete vertex normal based on A) volume gradient, $\nablar V$, mean curvature vector, $\vec{H}$, and Gaussian curvature vector, $\vec{K}$, and B) Schlafli vector, $H \vec{S}$}
    \label{SI_fig: ptwise direction}
\end{figure}

\subsection{Mesh mutation} 
\label{sec: mesh mutation}

Mesh mutation and refinement in combination with vertex shifting are the default methods to ensure that the mesh remains  well-conditioned and well-resolved during simulation. 
Mesh mutations include edge flipping, collapsing, and splitting, changes the connectivity of the mesh.
Vertex shifting moves the vertex to the barycenter of the fan neighborhood without changing the mesh topology (\cref{fig: energy and force}A). 
\mem3dg has a suite of possible criteria to initiate mesh mutation. 
Here we list the most important ones: 1) flip the edge of non-Delaunay diamond neighborhood (\cref{fig: energy and force}B), 2) collapse the shortest edge in a skinny triangle face, 3) split the edge with high (geodesic) curvature.
For additional details please refer to the software documentation.

For practical use, although mesh mutation introduces additional complexity in data writeout and computational costs associated with varying (usually growing) mesh size, it nevertheless provides a robust algorithm to ensure good mesh quality needed for valid discrete-smooth comparisons (\cref{sec: practical considerations for applying Mem3DG to biological problems}) in static frames.
For dynamical simulation, mesh mutations introduce an arbitrary interpolation of state variables, such as the position, velocity and protein density.
Rigorous study on how to interpolate these quantities to ensure the conservation of energy, momentum, and mass, remains to be done.
Similarly, the interpolation used in this study introduce discontinuities of curvature and can create jumps in forces;
This is particularly severe for terms with higher order derivatives such as the biharmonic term in bending force (\cref{eqn: discrete bending force}).

\subsection{Mesh regularization}
\label{sec: mesh regularization}

Mesh regularization can be used when mesh mutations are not desired. 
The regularization force consists of three weakly enforced constraining forces, including the edge (length), $\vbf^e$, face (area), $\vbf^f$, and conformality (angle), $\vbf^c$, regularization forces,
\begin{subequations}\label{eq: in-plane regularization}
 \begin{align}
    \vf^e_i & = -K_e  \nearbyEdges \frac{(l_{ij} - \bar{l}_{ij})}{\bar{l}_{ij}} \nablari l_{ij},  \\
    \vf^f_i &= -K_f \nearbyFaces \frac{(A_{ijk} - \bar{A}_{ijk})}{\bar{A}_{ijk}} \nablari A_{ijk}, \\
    \vf^c_i &= -K_c \nearbyEdges \frac{(\lambda_{ij} - 
    \bar{\lambda}_{ij})}{\bar{\lambda}_{ij}} \nablari \lambda_{ij},
\end{align} 
\end{subequations}
which are in the order of strongest to weakest.  
The length-cross-ratio, $\lambda_{ij} = {l_{il}l_{jk} } / {l_{ki}l_{jl}}$ is a metric of discrete conformality on triangulated mesh, where the indices is illustrated in \cref{fig: energy and force}A-B~\cite{Soliman2021}. 
Regularization forces require the input of a reference value for geometric measurements, $\bar{l}$, $\bar{A}$, and $\bar{\lambda}$, which can be derived from a well-conditioned reference mesh (usually the initial input mesh for the simulation).
The intensity of each regularization force is controlled with parameters $K_e$, $K_f$, and $K_c$.

For practical use, regularization constraints should be minimally imposed because of their impact on system dynamics.
In the worst case regularization constraints can prevent the optimizer from reaching an energy minima. 
Thus a good practice is to start a simulation in with no, conformality, face area, and finally edge length regularization, and subsequently raise the intensity/type of constraints based on the mesh quality desired.
We do not recommend imposing constraints stronger than the face areal constraints, $\vbf^f$.

\printbibliography
\end{document}